\documentclass[11pt]{article}
\usepackage{booktabs}
\usepackage[natbibapa]{apacite}
\makeatletter\renewcommand{\NAT@cmt}{; }\makeatother
\usepackage{graphicx}
\usepackage{amsmath}
\usepackage{amssymb}
\usepackage[hidelinks]{hyperref}
\usepackage{geometry}
\usepackage{cleveref}
\newcommand{\trait}[1]{\textsc{#1}}
\title{The rise and evolution of a referential code in
  populations of bee-like agents}
\author{Grzegorz Chrupała\thanks{Code, configurations, and result files to
    reproduce every figure and table in this paper are available at
    \url{https://github.com/gchrupala/bees}. AI coding agents, primarily
    Claude Code and Codex, were used in developing the codebase and as a
    writing aid in the preparation of this manuscript. The author reviewed all 
    content, and takes full responsibility for it.}\\
  Center for Cognitive Science and Artificial Intelligence\\
  Tilburg University\\
  \href{mailto:grzegorz@chrupala.me}{grzegorz@chrupala.me}}
\date{}

\begin{document}

\maketitle

\begin{abstract}
Communication typically relies on a shared code, and any change to it must be
coordinated between senders and receivers to avoid a breakdown of communication.
The honeybee waggle dance illustrates this problem: species with horizontal
combs point directly at a food source, while species with vertical combs cannot
point directly and instead reference the dance to gravity, decoded against the
position of the sun. We model the rise of the first of these codes and its
evolutionary transition to the second in populations of bee-like agents, with
selection acting at the level of colonies. In a horizontal-comb model, we find
that direct pointing evolves readily when food is moderately hard to find by
random search alone, whether because sites are few and large or many and small.
Communication fails to evolve when food is too sparse to spark dances or so
abundant that it is found without signaling. Adding an exogenous benefit for
vertical combs, we then find that the transition to the gravity-referenced code
is driven mainly by the magnitude of this benefit and by the mutation scale,
with the coupling between sender and receiver mutations playing a further role
when the mutation scale is low. Given a favorable confluence of these factors,
the transition proceeds reliably and without a breakdown of communication.
Outside that confluence it remains possible, though rarer, across a much wider
range of settings.
\end{abstract}

\section{Introduction}
\label{sec:intro}
Communication in animal species takes many forms, from the chemical
signals of ants \citep{jackson2006communication} to the symbolic,
compositional language of humans \citep{hockett1960origin}. Each of these communication systems has its own idiosyncratic
properties. What ecological and evolutionary pressures shaped these
systems has been of substantial interest to biologists and cognitive
scientists \citep{bradbury2011principles}.  In general, complex communication requires some kind of shared code
between senders and receivers, and the emergence of such a code
needs a degree of alignment between these two roles
\citep{lewis1969communication,skyrms2010signals}.
Likewise, any change or modification to the code needs to be
coordinated between senders and receivers if the breakdown of
communication is to be avoided.

In many cases it is possible to think of a plausible pathway for the
coordinated rise and development of a shared code
(\citealp{skyrms2010signals}, p.~29), but a verbally
formulated scenario leaves many details underspecified and makes it
hard to be precise about the specific conditions that make code change
trajectories feasible. For this reason computational models of the
evolution of communication systems can provide insights which are hard to
obtain by other means. Such models have been used extensively to study
the emergence of communication and language
\citep[][see \Cref{sec:related} for
details]{nowak1999evolution,smith2003iterated,wagner2003progress,skyrms2010signals,lazaridou2020emergent}. In
this paper we model a transition of a particularly intriguing
communicative system in invertebrates: the waggle dance of honey bees.

\begin{figure}
\centering
\includegraphics[width=0.8\textwidth]{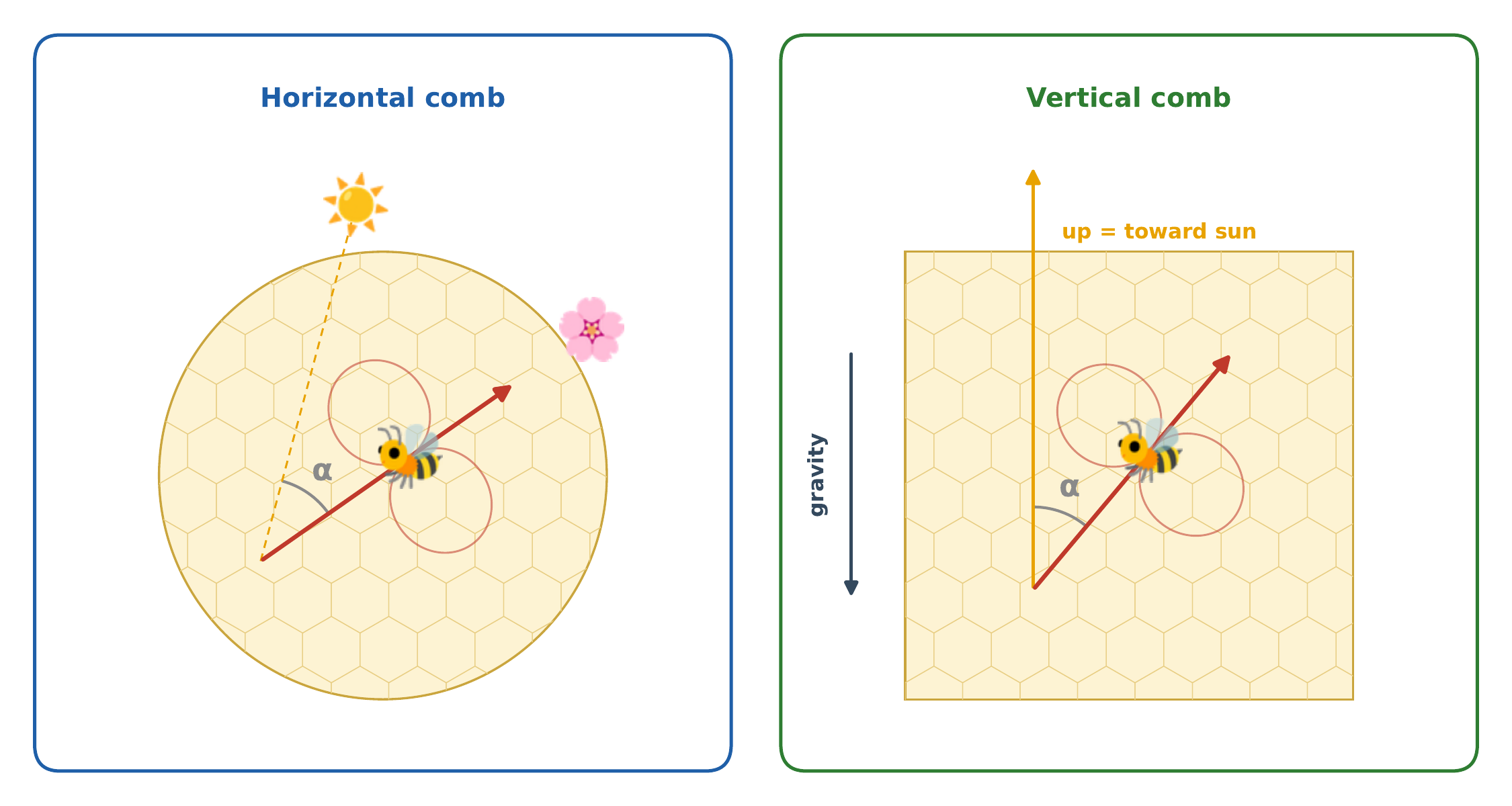}   
\caption{The two variants of the honeybee waggle dance. On a horizontal,
exposed comb (left) the bee aims the straight waggle run
directly along the true direction to the food, which it can identify as being 
at a particular azimuth relative to the sun.  
On a vertical comb (right) the
bee cannot directly point at the food. Instead, it uses gravity as a reference: 
it maps the direction \textit{up} to the position of the sun, 
and it rotates the dance axis away from vertical by an
angle $\alpha$ equal to the azimuth of the food site relative to the sun.
The figure-of-eight loops and the central waggle run are schematic and not to scale.}
\label{fig:waggle}  
\end{figure}

Communication is referential when signals stand for entities or states in the
world rather than merely expressing the signaler's internal state
\citep{macedonia1993variation,wheeler2012functionally}. The
communicative code of \emph{Apis} species \citep{vonFrisch1967DanceLanguage} is a unique
referential system where food sites discovered by foraging scouts are indicated
to other members of the colony by means of a figure-of-eight dance whose axis
encodes the direction towards the food source, and whose duration stands for the distance
to it. In this study we focus on the signaling of direction, as its evolution illustrates
some interesting questions about coordination between participants in a communicative
system.
Different species of \emph{Apis} vary in the specific dance variant used to
encode the direction of the food patch: in species with exposed horizontal combs
(\emph{Apis florea} and \emph{Apis andreniformis};
\citealp{hepburn2011biogeography}) the bees orient the axis of the
waggle dance directly towards the food patch
\citep[for \emph{A. florea} see][]{dyer1985mechanisms,dyer2002biology}.
In other members of the \emph{Apis}
genus, combs are vertical and in some species also hidden in cavities
\citep{hepburn2014nests}. Such
combs make direct pointing impossible, and these species use gravity as a
reference point, and map the azimuth of the food patch relative to the sun onto
the angle of the waggle dance with respect to the vertical
\citep{vonFrisch1967DanceLanguage,dyer2002biology}. \Cref{fig:waggle}
illustrates these two versions of the waggle dance.\footnote{This is a somewhat
simplified overview: in individual species there may be considerable behavioral
flexibility in how the dance is performed: e.g.\ in \textit{Apis mellifera}, sometimes
bees dance on a horizontal platform at the entrance to the beehive and in this case
they use direct pointing, even though this species generally employs the gravity
code \citep{Esch2012}.} 

The direct-pointing variant is iconic:
the form of the signal---the direction of the waggle axis---mirrors the direction
of the referent \citep{schlenker2026ancestral}. The ancestral, direct-pointing
dance has been characterized as 
an enactment of the foraging flight, in which the waggle run imitates the
departure  toward the food \citep{10.1242/jeb.142778}. The
gravity-referenced variant is more abstract, relying on an
indirect mapping of the sun's azimuth onto an angle measured from the vertical.
The waggle dance thus offers an opportunity to study  how a referential code
can shift from an iconic to a more abstract form, in a relatively simple system.

\begin{figure}
    \centering
    \includegraphics[width=0.8\textwidth]{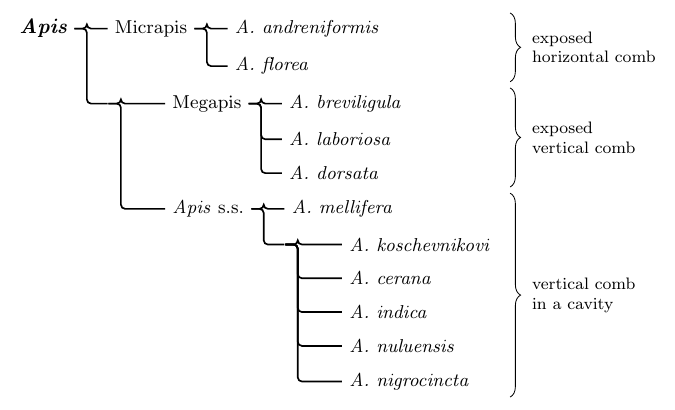}
    \caption{A simplified illustration of the \emph{Apis} phylogeny, with the three
      subgenera \emph{Micrapis}, \emph{Megapis} and \emph{Apis} sensu stricto. Adapted from
      \citet{https://doi.org/10.1111/j.1365-3113.2009.00504.x}}
    \label{fig:phylo}
\end{figure}

The distribution of comb orientation and waggle dance code throughout the \emph{Apis}
phylogeny (\Cref{fig:phylo}) suggests that the horizontal comb and direct
pointing are the ancestral state, extant in the stem \emph{Micrapis}
subgenus.\footnote{This conclusion has recently been contested:
  \citet{schlenker2026ancestral} reconstruct ancestral states over three
  \emph{Apis} phylogenies and find that the common ancestor was more
  likely than not to have danced vertically. We model the horizontal-to-vertical direction as the better 
  attested of two hypotheses rather than as established.}
Vertical
combs and the gravity-referenced code are derived traits, present in the other
two \emph{Apis} subgenera. Additionally, species within the \emph{Apis} sensu stricto
subgenus are cavity nesters. This pattern indicates that the
gravity code was linked to vertical combs and that cavity nesting is not
required for its emergence \citep{dyer1985mechanisms,10.1242/jeb.142778}.

Here we assume that such comparative differences in the dance reflect selection rather than
historical accident. This is established for the other component of the dance:
the code for communicating distance tracks
foraging range both across species and between populations of the same species, which supports
selection over neutral divergence \citep{kohl2020dialects}.

Nevertheless, the exact sequence and nature of evolutionary events that led to
the rise of the gravity-based waggle dance is not known in detail. It is also
unclear how the transition from one code to the other could happen gradually
while maintaining communicative success at every stage, or how this process
interacted with the change in comb orientation. Because the code is shared,
such a transition poses a coordination problem: a change adopted by senders
but not matched by receivers' interpretation, or the reverse, disrupts
communication, so intermediate stages risk a loss of communicative success.
A mechanistic account has been proposed at the neural level: the
central-complex circuitry that represents the bee's directional heading is
not tied to any particular spatial reference frame, so a gravity reference
could in principle substitute for a celestial one without a dedicated
switching mechanism \citep{10.1242/jeb.142778}. Such an account addresses how
a single brain can support either code, but not how a population of senders
and receivers stays coordinated as the shared code itself evolves, which
is the question we address.

The present study investigates this transition by means of a
computational model of evolving populations of bee-like agents. We
model selection at the level of colonies, and individual behavior at
the level of worker bees as colony members. We address two principal questions:
(i) under what ecological conditions a direct-pointing referential code
emerges and stabilizes, and (ii) which factors lead to a gradual
transition from this code to its gravity-referenced variant. Our aim is
not to reconstruct the precise historical sequence of these changes,
but to delineate the circumstances under which such a transition can
proceed without a breakdown of communication.

Our starting point is colonies with horizontal combs, which show some
heritable variation in traits controlling the inclination to direct
signaling and attending to signals. Here we show that:
communication via
direct pointing \emph{evolves readily when food is moderately hard to find by random
search alone}: i.e.\ when food sites are few in number but large, as well as when there 
are many small ones. It fails when food is too sparse to spark dances,
 or so abundant that it is found without signaling.

In the second stage we introduce extrinsic evolutionary pressure for
vertical combs, which partially degrades the direct-pointing code while
making the gravity reference available. We show that in this scenario the
transition towards the gravity-referenced code is driven mainly by the \emph{
mutation scale} and the \emph{magnitude of the exogenous benefit of the vertical comb.}
Additionally the \emph{strength of the coupling between the signaling behavior
of senders and receivers} also plays a role, especially in the low-mutation-scale
regime.

The pattern of results from these simulations indicates that a basic referential
code based on an iconic, direct-pointing signal evolves readily given favorable
ecological conditions. The transition from this basic ancestral system to the
abstract code based on gravity-referenced pointing is facilitated by the
confluence of several key ecological and evolutionary factors; once these hold,
however, the transition proceeds reliably and without a breakdown of
communication. At the same time, transition remains possible (if rare) in a
large region of less favorable parameter settings. 

\section{Related work}
\label{sec:related}
The current study builds on several distinct strands of work. Firstly, it 
draws on research into the emergence and evolution of communication in 
artificial multi-agent systems, viewed as an abstract computational problem.
 Secondly, because our work is inspired by and grounded in real honeybee 
 societies, it also builds extensively on models and simulations of bee 
 communication. These studies are methodologically diverse, employing 
 agent-based models, neural models, evolutionary computation, or combinations
 thereof.

 \subsection{Communication in agent societies}
A long line of computational work has studied how communication systems can
emerge and stabilize in populations of interacting agents. The signaling game
introduced by \citet{lewis1969communication,lewis1969coordination} provides the formal foundation for
much of this work, framing a convention of meaning as a solution to a recurring
coordination problem between a sender and a receiver. \citet{skyrms2010signals}
shows how such signaling conventions can arise without prior agreement through
evolutionary and reinforcement-learning dynamics. A distinct route emphasizes
cultural transmission: in the iterated-learning framework, linguistic structure
emerges as a code is repeatedly learned and reproduced across successive
generations of agents, without being built in \citep{smith2003iterated}. Related
formal work models how a protolanguage itself evolves under selection, showing
for instance how combining a small set of signals into words overcomes the error
limit that caps a simple signal repertoire \citep{nowak1999evolution}.
Agent-based models in this tradition have aimed to demonstrate that populations
playing repeated language games can self-organize shared, grounded vocabularies
and categories, a simulation tradition surveyed by \citet{wagner2003progress}.
More recently, deep multi-agent reinforcement learning has been used to show
that discrete communication protocols can emerge when agents are trained on
cooperative referential tasks \citep{lazaridou2017multi}. Subsequent work has
examined when such emergent codes become compositional and generalize to novel
meanings \citep{chaabouni2020compositionality}, and how communicative pressure
can drive neural agents towards attested language universals
\citep{lian2023communication}; see \citet{lazaridou2020emergent} for a review of
this line of research.

\subsection{Computational models of the waggle dance}
A venerable research tradition asks an ecological question: given that the dance
exists, \emph{when does it pay off}? The main tool here is the agent-based
foraging simulation. A colony of many simple forager agents follows fixed
behavioral rules. Dance recruitment is present as a switch, with recruits either
guided by a successful forager's spatial information or left to search on their
own, while the ecology (the density, quality, spread, and persistence of food
patches, and sometimes colony size) is varied across runs. The outcome of
interest is colony-level foraging performance: typically the net energy or
nectar collected. Importantly, these models represent neither evolution nor
within-lifetime learning: the communication strategy is fixed and the agents do
not adapt. What they model is the ecological cost-benefit accounting of a given
strategy: how a fixed colony fares, with recruitment versus without, in one
environment or another. Using such a model, \citet{dornhaus2006benefits} find
that recruitment benefits a colony mainly when resources are patchy, variable,
and hard to find, with the advantage depending on colony size, and
\citet{schurch2014dancing} show that long-term benefits of the dance can outweigh
its short-term costs. \citet{beekman2008pay} report the same dependence and add
that dance information can even reduce a colony's intake when patches are large
and close to the nest, where independent search is enough.
\citet{bailis2010positional} pose a more abstract version
of the same question, analyzing when positional communication works better than relying on
private information.  This body of work is the closest in spirit to our own
foraging payoff, but it holds the communication system constant and measures its
payoff in a fixed environment; it does not model the code itself as an evolving
trait.

More recently, the emergence itself of the bee communicative code has also been
investigated computationally. \citet{10.1371/journal.pcbi.1010487} treat
displaced communication as an evolutionary phenomenon: pairs of foraging agents,
each controlled by a continuous-time recurrent neural network, undergo
experimental evolution over tens of thousands of generations in a
one-dimensional circular arena where a receiver must reach one of several sites
a sender has perceived as holding food. Their central finding concerns how a
displaced signal arises. Although senders could in principle modulate  signal
amplitude, communication instead evolved from incidental, movement-derived cues
(the onset delay and duration of a sender's presence in the shared communication
zone), which were subsequently ritualized into signals. The more efficient
amplitude-based scheme evolved only when the cheaper cue was experimentally
suppressed. This model shares our evolutionary framing but selects at the level
of individual sender--receiver pairs rather than colonies, and it abstracts away
the geometry of the dance: the object of study is how any displaced signal
bootstraps from behavior, not how a specific directional code is structured.

\citet{siregar2026emergent} instead ask what internal representations and
channel properties allow a compositional, waggle-dance-like code to be
\emph{learned}. A single sender--receiver pair, implemented as graph neural
networks operating over map-like graphs of routes between the nest and candidate
sites, is trained by gradient descent on a referential game to transmit two
real-valued tokens identifying the food node. They ask whether the two tokens
factor into independent direction and distance components (compositionality),
how this depends on the overlap between the sender's and receiver's mental maps,
and whether constraining the channel to the kinematic range of a real dance
helps. They find that direction is readily encoded whereas direct distance is
not, that representational overlap is the main driver of compositionality, and
that channel constraints have only a small effect. This is a within-lifetime
study of representation and structure, with no population dynamics or selection.
In a similar vein, \citet{portegys2020morphognostic} has an individual agent
learn to both produce and follow dance-like movements that encode a nectar
location, again treating the dance as a code to be acquired within an
individual's lifetime rather than shaped by selection. The two differ in what is
left open: \citet{siregar2026emergent} let the code emerge from a referential
game and ask what structure results, whereas \citet{portegys2020morphognostic}
fixes the dance in advance and has a single agent learn to reproduce and respond
to it, studying acquisition of a known code rather than the emergence of its
structure.

\begin{table}
  \centering
  \small
  \begin{tabular}{@{}llll@{}}
    \toprule
    Approach & Code adapts by & Level & Question \\
    \midrule
    \multicolumn{4}{@{}l}{\emph{Communication in agent societies}}\\
    Signaling games and RL  & payoff dynamics       & population       & emergence \\
    Iterated learning       & transmission          & chain            & emergence \\
    Deep multi-agent RL     & learning              & pair             & emergence \\
    \addlinespace
    \multicolumn{4}{@{}l}{\emph{Models of the waggle dance}}\\
    Foraging simulations    & (fixed)               & colony           & payoff \\
    Evolved controllers     & evolution             & pair             & emergence \\
    Learned representations & learning              & individual/pair  & learnability \\
    \midrule
    \textbf{This work}      & evolution             & colony           & transition \\
    \bottomrule
  \end{tabular}
  \caption{Computational approaches to how a communicative code arises or
    changes, and where the present model sits among them. The columns give how
    the code adapts (if at all), the level it applies to, and the question each
    line of work asks. Rows are grouped into general models of communication and
    models specific to the waggle dance; each row is discussed and cited in the text.}
  \label{tab:routes}
\end{table}

Our study occupies a different point in this space (\Cref{tab:routes}). We model
the emergence and in particular the evolutionary \emph{transition} between two
forms of a referential code at the level of colonies, using a small set of
interpretable, heritable traits rather than opaque neural controllers. This lets
us treat costs, benefits, and comb geometry as explicit parameters and ask which
ecological and evolutionary conditions drive the shift from iconic direct
pointing to an abstract gravity-referenced code: a question about the conditions
and dynamics of a communicative transition that none of the prior work
addresses. 

\section{Method}
\label{sec:method}
Our model is inspired by bee colonies and their waggle dance, but it is
simplified in many important ways. Our goal is not to model bee biology in
detail, but to investigate scenarios for the rise and evolution of
referential communicative codes. We adopt the case of the waggle dance as a way
to ground the modeling in the real world and ensuring that it is evolutionarily
plausible.

We model evolution at the level of colonies, and behavior at the level of
individual bee-like agents. This treatment is motivated by the reproductive and social
organization of real bee colonies, where workers are highly related and
sterile, and thus selection acts effectively on heritable colony-level traits
rather than on individual workers---this is the standard kin-selection rationale
for treating the honeybee colony as a superorganism
\citep{hamilton1964genetical,seeley1989superorganism}. We do not model sexual
reproduction: agent colonies reproduce without genetic input from other colonies. 
Colonies have a number of heritable
traits which are transmitted to offspring colonies subject to mutation, and
individual agents express these traits with some non-heritable individual
variation. In the interest of simplicity and focus, we model the waggle dance as only signaling 
the direction of a food site: our agents do not indicate distance.

\subsection{Colonies, workers, and traits}
\label{sec:traits}
We keep two kinds of quantity visually distinct throughout: a colony's heritable
traits, which evolve under mutation and selection and whose names we set in
\trait{small capitals}, and the fixed or swept simulation parameters of the model,
which are chosen by the experimenter, referred to by their mathematical symbols,
and listed in \Cref{tab:fixed-params,tab:varied-params}. A colony is described
by a small set of heritable, real-valued traits, listed in full in
\Cref{tab:traits}. The prose here introduces the traits needed for the basic
horizontal-comb model (\Cref{sec:horizontal}); the additional traits and
geometry required for tilted combs are developed in \Cref{sec:vertical}.

\begin{table}
  \centering
  \small
  \begin{tabular}{@{}cll p{0.42\textwidth}@{}}
    \toprule
    Symbol & Name & Range & Role \\
    \midrule
    \multicolumn{4}{@{}l}{\emph{Behavioral (horizontal-comb model, \Cref{sec:horizontal})}}\\
    $b$      & \trait{directional bias}    & $[0,1]$ & sharpness of a dance about the food direction \\
    $a$      & \trait{receiver attention}  & $[0,1]$ & probability of following a dance rather than searching at random \\
    $p$      & \trait{dance propensity}    & $[0,1]$ & readiness to recruit others after a successful foraging attempt \\
    $\ell$   & \trait{foray distance}      & $[0,D]$ & mean outbound distance of a foray \\
    \addlinespace
    \multicolumn{4}{@{}l}{\emph{Nest and code choice (vertical transition, \Cref{sec:vertical})}}\\
    $\gamma$ & \trait{comb tilt}              & $[0,1]$ & comb inclination, $0$ horizontal to $1$ vertical \\
    $\phi$   & \trait{comb orientation}       & $[0,P)$ & compass heading the comb faces \\
    $t_s$    & \trait{sender transposition}   & $[0,1]$ & weight for the gravity code when producing a dance \\
    $t_r$    & \trait{receiver transposition} & $[0,1]$ & weight for the gravity code when interpreting a dance \\
    \bottomrule
  \end{tabular}
  \caption{The heritable, colony-level traits of the model. The first group
    governs behavior on a horizontal comb (\Cref{sec:traits,sec:horizontal}); the
    second adds the nest geometry and code-choice traits used in the vertical
    transition (\Cref{sec:vertical}). Workers express $b$, $a$, $p$, $t_s$, and
    $t_r$ with individual variation (\Cref{sec:traits}); $\ell$ varies 
    through the per-foray draw (\Cref{sec:environment}), and the nest traits
    $\gamma$ and $\phi$ are shared by all of a colony's workers. Here $D$ is the
    maximum search distance and $P$ the orientation period.}
  \label{tab:traits}
\end{table}

Four traits govern the behavior of a colony's workers. The \trait{directional
bias} $b$ controls how precisely a dance points; the \trait{receiver attention}
$a$ is the propensity to attend to a dance rather than to search at random; the
\trait{dance propensity} $p$ controls how readily a successful scout recruits
others to the patch it just found (\Cref{sec:foraging}); and the \trait{foray
distance} $\ell$ sets the mean outbound distance a worker travels on a foray. The first three are bounded to $[0,1]$ and the \trait{foray distance} to $[0,D]$,
where $D$ is the maximum search distance, a fixed model parameter (set to
$D = 6$ km in all our experiments).

Workers express the colony traits with non-heritable individual variation. A
worker $i$ realizes each trait $x \in \{b, a, p\}$ as
\begin{equation}
  x_i = \mathrm{clip}\bigl(x + \varepsilon_i\bigr), \qquad
  \varepsilon_i \sim \mathcal{N}(0, \eta^2),
\end{equation}
where $\eta$ is a fixed worker-variation scale and $\mathrm{clip}$ returns the
value to its admissible range. The \trait{foray distance} $\ell$ carries no separate
worker jitter: within-colony variation in how far workers travel arises instead
from the per-foray draw described in \Cref{sec:environment}. A colony thus
behaves as a noisy ensemble of workers sharing the same underlying genotype.

\subsection{Producing and interpreting signals}
\label{sec:signals}
On a horizontal comb the dance points directly at the food: this is the iconic,
direct-pointing code. A worker that has just foraged successfully at azimuth $\alpha$
produces a dance whose intended direction is $\alpha$ itself. The emitted signal is
drawn from a von Mises distribution centered on $\alpha$ with concentration
$\kappa = b_i\,\kappa_{\max}$, plus a small wrapped Gaussian dance noise. The von
Mises distribution is the circular analogue of a Gaussian: a signal $\psi$
centered on $\mu$ with concentration $\kappa$ has density
\begin{equation}
  f(\psi;\mu,\kappa) = \frac{\exp\bigl(\kappa\cos(\psi-\mu)\bigr)}{2\pi I_0(\kappa)},
\end{equation}
where $I_0$ is the modified Bessel function of order zero and $\kappa$ plays the
role of an inverse variance; here $\mu = \alpha$ and $\kappa = b_i\,\kappa_{\max}$.
The \trait{directional bias} $b$ therefore
controls the sharpness of the dance: a low bias leads to a near-uniform,
uninformative signal, while a high bias gives rise to a tightly concentrated one that
reliably indicates $\alpha$.

A worker that follows a dance reads its direction as a search direction, subject
to a small added interpretation noise. Communication is successful when this
recovered direction is close enough to a food site to lead the follower to it.

\subsection{Foraging environment}
\label{sec:environment}
\begin{figure}
  \centering
    \includegraphics[width=0.8\textwidth]{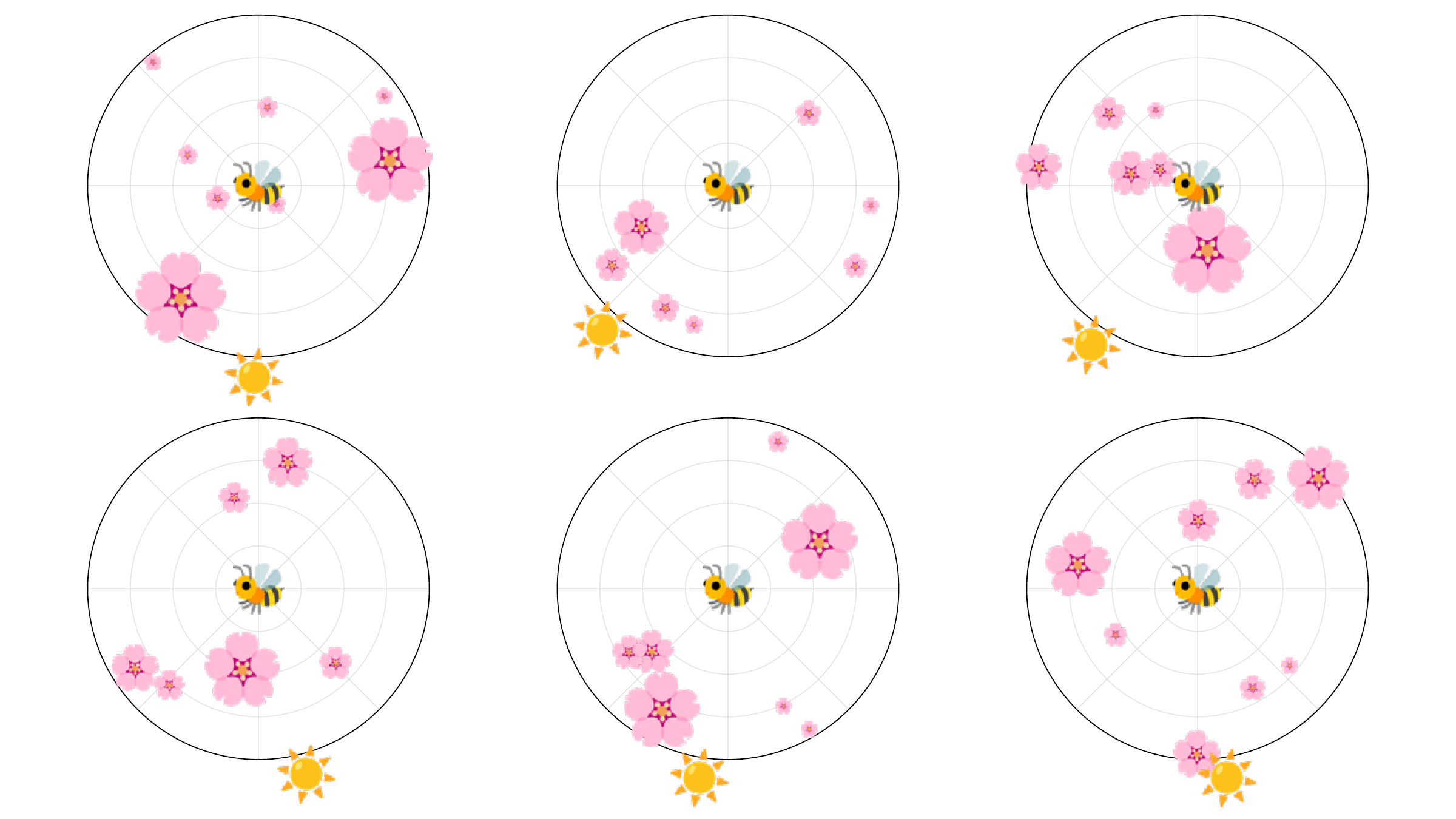}
    \caption{Six samples of the foraging environment. Each panel
      shows the colony (bee) at the centre, the food patches (flowers) at their
      drawn azimuths and distances, and the sun, which marks the external
      reference for the gravity-based code. Each patch is a physical disk, and
      the flower size scales with the sampled radius of the patch; radial rings mark
      distance from the colony in kilometers. The panels are drawn at
      representative values chosen to illustrate the geometry: the number of
      patches per episode is Poisson with mean $6$, so panels differ in how many
      patches they contain, the median radius is $150$ m (lognormal, log-sd
      $0.6$), and patches lie out to a maximum distance of $6$ km. All three sit
      within the ranges the experiments explore
      (\Cref{sec:setup-horizontal,sec:setup-transition}).}
    \label{fig:environment}
\end{figure}

A colony is evaluated over a fixed number of independent foraging episodes, and
its fitness is averaged over them (\Cref{sec:foraging}). Each episode draws a
fresh environment (\Cref{fig:environment}), with patch radii in meters and
distances from the nest in kilometers. The number of food sites is drawn from a
Poisson distribution with mean $n$, so episodes vary in how much food they offer.
 Each site is a physical circular disk whose radius is drawn independently
from a lognormal distribution with median $\mu_r$ and log-scale spread
$\sigma_r$. It is placed at an azimuth drawn uniformly around the colony,
with its distance chosen so that sites are uniformly dense per unit area
between $d_{\min}$ and $d_{\max}$. Concretely, the squared distance is drawn
uniformly between $d_{\min}^2$ and $d_{\max}^2$. This places most patches in the
outer part of the range, the sparse-landscape regime in which distant patches are
thought to be exploitable only by colonies that recruit
\citep{beekman2000longrange}. A
forager finds the patch when its straight outbound path crosses the disk, so a
fixed patch subtends a smaller angle when farther from the nest, and distant
food demands more accurate dances. Each patch has value $v$ per visit and a finite
capacity limiting how many foragers it can supply before being exhausted;
capacity scales with the patch area, growing with the square of its radius, so
that larger patches hold proportionally more food while the per-visit value
stays fixed. Throughout, $v = 1$, the log-scale spread is $\sigma_r = 0.6$, and
sites lie no closer than $d_{\min} = 0.75$ km; the median radius $\mu_r$, the
capacity at that radius, and the maximum distance $d_{\max}$ vary by experiment
(\Cref{tab:varied-params}).

The \trait{foray distance} trait $\ell$ (\Cref{sec:traits}) sets how far workers travel.
Each individual foray draws its own outbound length from a Gamma distribution
with mean $\ell$ and a fixed shape, capped at the maximum search distance $D$. The
per-foray draw is the source of within-colony variation in travel distance and
gives a realistic spread of short and long forays around the colony mean.

Each episode also fixes the position of the sun, which serves as the external
reference for the gravity-based code introduced in \Cref{sec:vertical}. The
sun's azimuth is drawn uniformly from an arc of width $\Delta_{\mathrm{sun}}$
centered on a fixed heading $\mu_{\mathrm{sun}}$ (in our experiments
$\mu_{\mathrm{sun}} = \pi/2$ and $\Delta_{\mathrm{sun}} = \pi$), so it is constant
within an episode but varies between episodes. We explain the role of this
variable sun azimuth in \Cref{sec:vertical}.

\subsection{Foraging and colony payoff}
\label{sec:foraging}
Within each episode (\Cref{sec:environment}) the colony makes a sequence of
foraging attempts. On each attempt a worker is drawn at random; if at least one
dance has already been performed in the current episode,
the worker follows a randomly chosen dance with probability equal to its
attention $a_i$, reading it out as a search direction, and otherwise searches in a
uniformly random direction. The worker then sets out along that direction on a
foray whose length is drawn as in \Cref{sec:environment} and enters the first
food patch on its path, among patches that still have remaining
capacity and lie within the foray length; if no patch is reached the attempt
fails. The fraction of foraging attempts that reach a patch is their
\emph{foraging success}.

Payoff accrues per episode. A successful foraging attempt yields the patch value $v$, minus
a travel cost $c_{\mathrm{travel}}$ per unit of patch distance, and may prompt the worker to
perform a dance advertising the patch to later foragers. Recruitment is itself an
evolved decision: a successful scout dances with probability $1 - (1 - p_i)^{C}$,
where $p_i$ is its \trait{dance propensity} and $C$ is the remaining capacity of the patch
after the visit, so an exhausted patch never leads to a dance and a scarcely
provisioned one rarely does. A dance, when performed, costs
$c_{\mathrm{base}} + c_{\mathrm{cue}}\,b_i$, growing with the precision of the
signal. A failed attempt costs travel proportional to the foray length it
travelled, and each act of attending to a dance incurs an attention cost $c_{\mathrm{att}}$.
Collecting these terms, an episode with successful attempts $S$ (each at distance
$r_k$), the subset $S_d \subseteq S$ of successes that produced a dance (each by a
worker with bias $b_k$), failed attempts $F$ (each with foray length $\ell_j$),
and $n_a$ dance-following attempts yields raw payoff
\begin{equation}
  \pi_{\mathrm{raw}} = \sum_{k \in S}\bigl(v - c_{\mathrm{travel}}\,r_k\bigr)
    - \sum_{k \in S_d}\bigl(c_{\mathrm{base}} + c_{\mathrm{cue}}\,b_k\bigr)
    - \sum_{j \in F} c_{\mathrm{travel}}\,\ell_j
    - c_{\mathrm{att}}\,n_a,
\end{equation}
where the dance-cost sum runs over $S_d$ because a success produces a dance only
when the recruitment decision is positive.
The scaled episode payoff is $\pi = (1 + B\gamma)\,\pi_{\mathrm{raw}}$, where
$\gamma$ is the \trait{comb tilt} and $B \ge 0$ the vertical-comb benefit; in the
horizontal model $\gamma = 0$, so the scaling is inert. We motivate its form in
\Cref{sec:transition}. A colony's fitness is its mean episode payoff (floored at a
small positive value).
For diagnostic purposes we also record a \emph{recruitment advantage}: the
difference between the foraging success of workers that followed a dance and that
of contemporaneous workers that searched at random while dances were available. A
positive recruitment advantage indicates that the code is functioning, since
following a dance then beats searching blindly.

\subsection{Evolutionary dynamics}
\label{sec:evolution}
Evolution proceeds in non-overlapping generations over a fixed-size population of
colonies; there is no sexual reproduction. In each generation every colony is
evaluated, and the next generation is produced by repeated 
selection of parent colonies with a probability proportional to fitness. 
A selected colony's traits are copied subject to
mutation. Each trait is perturbed by an independent Gaussian step and then clipped to its
range. The standard deviation of the step is a fixed fraction $\sigma_m$ of that
range: $\sigma_m$ for the traits on the unit interval, and $\sigma_m D$ for the
\trait{foray distance}, whose range is $[0,D]$. The workers of each offspring colony are then regenerated from its traits
as in \Cref{sec:traits}. The mutation scale $\sigma_m$ thus sets the pace of
evolutionary change.

\subsection{Horizontal comb}
\label{sec:horizontal}
The horizontal-comb model is exactly as described above: the comb is flat
and the only code available is direct pointing. We use it to
investigate the conditions under which a referential code emerges and stabilizes,
starting from an initial population with little communicative ability (low
\trait{directional bias} and low attention), so that early foragers signal weakly and
rarely attend to one another.

\subsection{Modeling code transition}
\label{sec:transition}
We ask how a colony can move from the direct-pointing code, available on a
horizontal comb, to the gravity-referenced code used on a vertical one. We do not
model why combs come to hang vertically; instead we impose an exogenous benefit
for vertical building through the payoff scaling $(1 + B\gamma)$
(\Cref{sec:foraging}), which rewards greater \trait{comb tilt} on its own, and ask
whether communication can survive as the comb tilts under that pressure. The
scaling is multiplicative because the advantages usually attributed to a vertical
comb are architectural:\footnote{A vertical comb carries cells on both faces of a
single sheet, whereas a horizontal one can open its cells only upward; and a comb
attached along its top edge is loaded in its own plane rather than in bending,
which wax tolerates better \citep{seeley1976nest,hepburn2014nests}.} they improve the return on what a colony
already forages, rather than providing a bonus that arrives regardless of
foraging success. This form couples the two pressures: comb tilt pays off
most for a colony that already forages well, which is exactly the colony
with a working direct code to lose. Because
\trait{comb tilt} is itself a trait that starts flat and evolves by small
mutational steps, a lineage passes through intermediate, partly tilted combs
rather than jumping from horizontal to vertical.

Those intermediate tilts are what make the transition non-trivial. As the comb
tilts, direct pointing degrades: projecting the food direction onto the sloping
surface foreshortens it in a bearing-dependent fashion, leaving less of the
direction in the surface to carry the signal. How faithfully the heading is
recovered depends on whether the receiver corrects for comb orientation. We thus
model two decoders (\Cref{sec:vertical}): a more biologically plausible
\emph{flatten} decoder which applies no such correction and is therefore biased
on a tilted comb, and an idealized \emph{unproject} decoder that corrects
exactly and serves as a bias-free reference against which to read the results of
\emph{flatten}.

Meanwhile, the growing tilt makes the
gravity-referenced code increasingly reliable, so a colony can in principle
switch codes as direct pointing weakens. We again model this switch as gradual,
through a blend of the two codes (\Cref{sec:vertical}).

\subsection{Dancing on a tilted comb}
\label{sec:vertical}
The tilted-comb model extends the horizontal one with the machinery for comb
tilt and the gravity-referenced code. It introduces two further
colony-level traits describing the nest: the \trait{comb tilt} $\gamma \in [0,1]$,
where $\gamma = 0$ is a horizontal comb and $\gamma = 1$ a vertical one, and the
\trait{comb orientation} $\phi$, the compass heading the comb faces. Like the
behavioral traits these are heritable and mutate by an independent Gaussian step
each generation, but they are
properties of the nest and so are shared by all of a colony's workers rather than
being expressed with individual variation. Because the orientation $\phi$ is an angle
defined on a circle of period $P$, its mutation step has standard deviation
$\sigma_m P$ rather than $\sigma_m$.

\paragraph{Comb geometry.}
The comb is a flat surface on which dances are performed, and its tilt determines
which reference directions a dance can use. We represent the comb by its unit
normal, the direction perpendicular to its face. When the comb is horizontal
this normal points straight up; as the comb tilts, the normal swings down toward
the horizon in the compass direction $\phi$, until on a vertical comb it lies
flat. Writing the tilt as a physical angle $\theta = \gamma\,\pi/2$ (so
$\theta = 0$ for a horizontal comb and $\theta = \pi/2$ for a vertical one), the
normal has eastward, northward, and upward components
\begin{equation}
  \mathbf{n} = \bigl(\,
    \underbrace{\sin\theta\cos\phi}_{\text{east}},\;
    \underbrace{\sin\theta\sin\phi}_{\text{north}},\;
    \underbrace{\cos\theta}_{\text{up}}
  \,\bigr).
\end{equation}
At $\theta = 0$ this reduces to $(0,0,1)$, pointing straight up, and at
$\theta = \pi/2$ it lies in the horizontal plane pointing along $\phi$.

\begin{figure}
  \centering
  \includegraphics[width=0.6\textwidth]{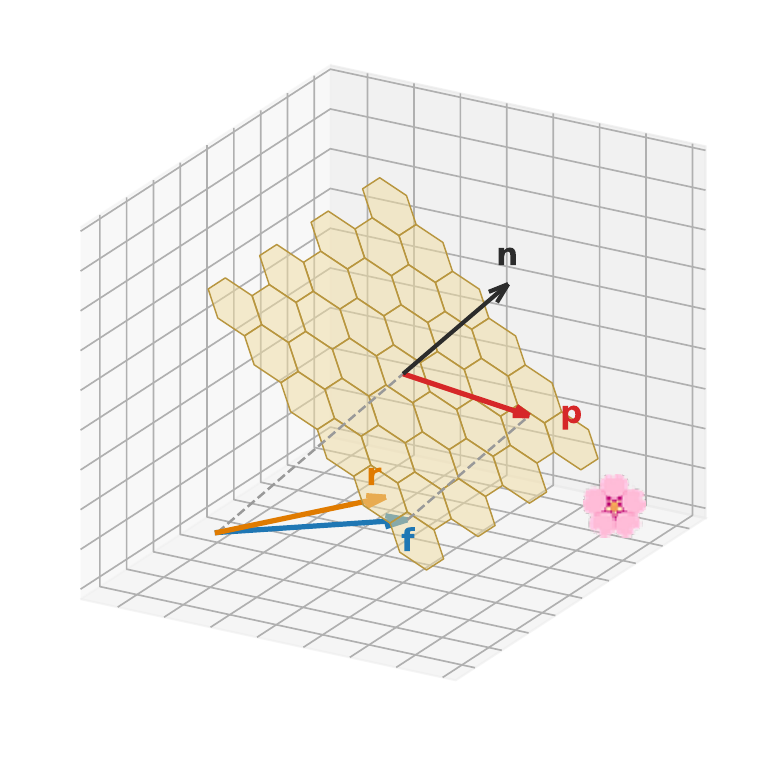}
  \caption{The projection of the direct code and the bias of the flatten decode. The
    tilted hexagonal comb and the world-horizontal plane (the grid) share the
    east--north position of the hive. The food direction $\mathbf{f}$ (blue) lies
    in the world plane; its orthogonal projection onto the comb, $\mathbf{p}$
    (red), found by carrying its start and end points along the comb
    normal $\mathbf{n}$ (black) until they meet the surface, shown by the
    dashed rays, has in-surface angle $\delta_\mathrm{dir}$, the encoded
    signal. Decoding that signal back to a world heading, the flatten variant
    recovers $\mathbf{r}$ (orange), which drifts off the true direction by a
    systematic angle on any tilted comb, whereas the unproject variant would
    recover $\mathbf{f}$ (blue) exactly. The flower marks the food site.}
  \label{fig:direct-decode}
\end{figure}

\paragraph{The two codes.}
Both codes express the food direction as an angle in the comb surface; they
differ in the reference the angle is measured from and in how much of the
direction the surface can carry. In the \emph{direct} (iconic) code the worker
projects the world direction of the food, $\mathbf{f} = (\cos\alpha, \sin\alpha,
0)$, orthogonally onto the comb and dances along the resulting in-surface vector
$\mathbf{p}$ (\Cref{fig:direct-decode}).

In the orthonormal surface axes $\mathbf{e}_1, \mathbf{e}_2$ of the comb, this
projection has coordinates $(p_1, p_2) = (\mathbf{e}_1\cdot\mathbf{f},\;
\mathbf{e}_2\cdot\mathbf{f})$, and the encoded signal is its angle
$\delta_{\mathrm{dir}} = \operatorname{atan2}(p_2,p_1)$. The \emph{strength} of the code is how much of the unit direction survives the projection,
$s_{\mathrm{dir}} = \lVert\mathbf{p}\rVert$. What the projection discards is the
part of $\mathbf{f}$ along the comb normal, $\mathbf{f}\cdot\mathbf{n} =
\sin\theta\cos(\alpha-\phi)$, leaving
\begin{equation}
  s_{\mathrm{dir}} = \sqrt{1 - \sin^2\theta\,\cos^2(\alpha-\phi)}.
\end{equation}
On a horizontal comb $s_{\mathrm{dir}} = 1$; that is the direction lies entirely within
the surface and is represented faithfully. As the comb tilts $s_{\mathrm{dir}}$
shrinks, reaching its minimum $\cos\theta$ for the food site in the direction the comb
faces ($\alpha = \phi$). Only on a fully vertical comb does this minimum fall to
zero: the facing direction then projects to nothing and its heading cannot be
recovered, while every other direction stays informative.

In the \emph{gravity-referenced} code the worker instead projects the vertical
direction $\mathbf{g} = (0,0,1)$ onto the comb to obtain an in-surface reference,
and encodes the food azimuth, measured relative to the sun, as an angle from that
reference. Its strength depends only on tilt, not on the bearing to the food site,
\begin{equation}
  s_{\mathrm{grav}} = \sin\theta = \sin(\gamma\,\pi/2),
\end{equation}
behaving oppositely to $s_{\mathrm{dir}}$: it is zero on a horizontal comb, where
gravity is perpendicular to the surface and gives no in-surface reference, and
grows to $1$ as the comb becomes vertical. The two trade off: a horizontal comb
supports only direct pointing, a vertical comb makes the gravity reference
available for every food direction, and at intermediate tilts both carry
information.

When code strength is below $1$, the angle a worker computes for either
code---whether producing or interpreting a signal---is independently degraded,
modeled as a circular interpolation between the true angle and a uniformly
random one, with weight equal to the strength.

\paragraph{Decoding the direct code.}
The direct encoding is linear in the horizontal food vector:
$(p_1,p_2)^\top = M(\cos\alpha,\sin\alpha)^\top$, where $M$ is the $2\times 2$
matrix whose rows are the east--north parts of $\mathbf{e}_1,\mathbf{e}_2$, with
$\det M = \cos\theta$. Recovering the world heading exactly requires inverting
this map, and we model two receivers that differ in whether they do so
(\Cref{fig:direct-decode}). The inversion asks a lot of a receiver: it must know
\trait{comb tilt} and \trait{comb orientation}, and apply a correction that
varies with the direction danced. The uncorrected reading asks nothing, since the
dance is a direction in space and the food is on the ground: a receiver that
flies along the compass bearing of the dance drops the vertical component simply by
flying. That reading is also exactly right on a horizontal comb (the ancestral
state), which is why we regard it as the more plausible of the two: it leaves the
ancestral rule unchanged, whereas the correction would have had no function to
select it before combs tilted.
\begin{description}
  \item[Unproject] The receiver applies $M^{-1}$, exactly inverting the
    encoding; since $\det M = \cos\theta > 0$ for $\theta < \pi/2$, the true
    heading is recovered without bias.
  \item[Flatten] The receiver reads the dance as a direction in space and drops
    its vertical component, taking the compass bearing of what is left. Where
    the sender projected the food direction onto the comb, the receiver projects
    the danced direction back onto the horizontal plane: a second projection
    rather than an undoing of the first, so the recovered heading $\mathbf{r}$
    is systematically biased on a tilted comb.\footnote{Formally, the receiver
    lifts $\delta$ to the in-surface vector $\cos\delta\,\mathbf{e}_1 +
    \sin\delta\,\mathbf{e}_2$ and keeps its east--north part, which applies
    $M^\top$ rather than $M^{-1}$. On a flat comb $M$ is a rotation, so the two
    coincide and the decodes are identical; they differ only once the comb
    tilts.}
\end{description}

The gravity code, by contrast, is always decoded by exact inversion:
the projected gravity reference is available to any receiver, so the sun-relative
food azimuth is read back directly.

\paragraph{Transposition and blending the codes.}
Which code a worker uses is governed by two further heritable behavioral traits,
the \trait{sender transposition} $t_s$ and \trait{receiver transposition} $t_r$
(both in $[0,1]$ and expressed with individual variation as in
\Cref{sec:traits}). These set how much weight a worker places on the
gravity-referenced code rather than the direct code when, respectively,
producing and interpreting a dance.\footnote{The alternative is a binary trait,
with a worker either pointing directly or referring to gravity. We use graded
weights because they let \trait{transposition} change by the same small
mutational steps as every other trait here. Additionally, there
is some biological support for blending: bees put into conflict between a light
reference and gravity dance at compromise angles \citep{edrich1977interaction}.
Blending is nonetheless the more permissive assumption: a binary trait would
make mismatched senders and receivers mutually unintelligible, so a population
would pay a coordination cost at intermediate frequencies that a blending one
avoids.}

Both roles blend the codes by the same rule. Given the angles
$\delta_{\mathrm{dir}}$ and $\delta_{\mathrm{grav}}$ the two codes assign, and a
transposition $t$, each angle is turned into a unit vector, scaled by its weight,
the two vectors are summed, and the blend is the heading of the resultant:
\begin{equation}
  \operatorname{blend}(\delta_{\mathrm{dir}}, \delta_{\mathrm{grav}};\, t)
    = \operatorname{atan2}\Bigl(\textstyle\sum_k w_k \sin\delta_k,\;
      \sum_k w_k \cos\delta_k\Bigr),
  \qquad k \in \{\mathrm{dir}, \mathrm{grav}\},
\end{equation}
where the weights combine the transposition with the geometric strength of
each channel,
\begin{equation}
  w_{\mathrm{dir}} = (1 - t)\,s_{\mathrm{dir}}, \qquad
  w_{\mathrm{grav}} = t\,s_{\mathrm{grav}}.
\end{equation}
A sender blends the in-surface angles the two codes assign to the food just
visited, giving the intended dance direction
$\delta = \operatorname{blend}(\delta_{\mathrm{dir}}, \delta_{\mathrm{grav}};\,
t_{s,i})$, and emits a signal from a von Mises distribution about $\delta$ as in
\Cref{sec:signals}. A receiver decodes the observed signal under both codes and
blends the two recovered \emph{world} directions the same way, with $t_{r,i}$ in
place of $t_{s,i}$.

A worker with $t_i = 0$ therefore relies purely on direct pointing and one with
$t_i = 1$ purely on the gravity reference, while the strength factors ensure that
a code with no geometric support (such as the gravity code on a horizontal comb)
contributes nothing regardless of $t$. The geometric strength thus enters in two
distinct roles: it degrades the world direction each channel recovers,
interpolating it toward a uniformly random heading as described above, and it
scales the weight of that channel in the blend, so a channel with weak geometric
support is at once less reliable and less relied upon.

Communication is accurate to the extent that senders and receivers weight the two
codes alike: the further $t_{r,i}$ sits from $t_{s,i}$, the more the receiver
reads the dance under a mix of references the sender did not use, and the larger
the resulting directional error. Alignment only matters where the codes disagree,
however, so on a horizontal comb, where $s_{\mathrm{grav}} = 0$ collapses both
blends onto the direct code, $t_r$ is free to drift.

The per-episode sun azimuth (\Cref{sec:environment}) matters through this
blending. Within an episode the sun is shared by every worker, so when a dancer
and follower both use the gravity code it enters encoding and decoding with
opposite sign and cancels, and where it falls does not matter. Because it varies
across episodes, however, a colony cannot treat it as a fixed offset: a strategy
that only partially commits to the gravity code leaves an uncancelled sun term,
so the shifting reference selects for genuine transposition rather than a stable
blend of the two codes.

\paragraph{Coupling of sender and receiver.}
A change of code pays off only to the extent that it is matched on the other
side, so the two traits may plausibly not mutate independently. We therefore
allow the mutations of $t_s$ and $t_r$ to be correlated, treating the strength of
the coupling as a parameter. Their mutational steps are drawn as a
correlated Gaussian pair whose sender--receiver coupling is the
correlation coefficient $\rho$,
\begin{equation}
  \Delta t_s = z_1, \qquad
  \Delta t_r = \rho\,z_1 + \sqrt{1-\rho^2}\;z_2, \qquad
  z_1, z_2 \sim \mathcal{N}(0, \sigma_m^2),
\end{equation}
which leaves the marginal mutation scale of each trait unchanged while tuning how
tightly the two co-vary. When $\rho$ is high, a mutation that pushes senders
toward the gravity code tends to push receivers by the same amount, so the two
sides of the code can shift in a coordinated fashion rather than drifting apart;
when $\rho = 0$ they evolve independently. The \trait{comb tilt} $\gamma$, held fixed at
$0$ in the horizontal model, is free to mutate in this version.

\section{Experimental setup}
We study the model of \Cref{sec:method} in two stages that mirror the two
questions of \Cref{sec:intro}. The first stage keeps the comb horizontal and asks
under what ecological conditions the direct-pointing code is worth maintaining
(\Cref{sec:setup-horizontal}). The second stage releases the comb and the
\trait{transposition} traits and asks under what conditions a population can migrate from
the direct to the gravity-referenced code as an extrinsic pressure tilts the comb
(\Cref{sec:setup-transition}).

\subsection{Model parameters}
\label{sec:setup-fixed}
A core set of parameters is held fixed throughout, and only the ecological and
evolutionary parameters named in \Cref{tab:varied-params} are varied. 
\Cref{tab:fixed-params} lists the fixed values. We do not attempt to model the 
physiology and ecology of a particular bee species in detail; however we do try to
make sure the fixed parameters values are reasonably plausible in order to make the results
 interpretable in biological terms. 

The population contains $60$ colonies of $80$ workers each
(\citealp{visscher1982foraging} estimate $\approx 50$ feral colonies share one
colony's foraging range), evolving in non-overlapping generations; each colony is evaluated over
$50$ independent foraging episodes of $12$ foraging attempts, each attempt made
by a single worker drawn at random (the worker pool is therefore a sample of the
colony's trait distribution rather than a workforce). Workers
express colony traits with variation scale $\eta$ (\Cref{sec:traits}); a dance
indicates its direction with limited angular precision, bounded by
$\kappa_{\max}$ and reached only at maximum \trait{directional bias}; and every
dance carries both production and interpretation noise (\Cref{sec:signals}).
All lengths follow a
fixed convention (distances in kilometers, patch radii in meters): the maximum
search distance is $D = 6$ km, the $95$th percentile of foraging distance
recorded by \citet{visscher1982foraging}, and food patches lie at distances in
$[0.75, d_{\max}]$ km. Food patches are physical disks
whose radius is drawn from a lognormal and whose capacity scales with patch area,
and the number of patches per episode is Poisson (\Cref{sec:environment}). Each foray draws its
outbound distance from a Gamma of fixed shape with the colony's evolved mean.
Foraging carries a per-distance travel cost, a dance cost
$c_{\mathrm{base}} + c_{\mathrm{cue}}\,b_i$ that is purely bias-dependent
($c_{\mathrm{base}} = 0$), and an attention cost $c_{\mathrm{att}}$
(\Cref{sec:foraging}). The sun is drawn from an arc of width $\pi$ centered on
$\pi/2$, and the \trait{comb orientation} is treated as axially symmetric (period $\pi$).
Colonies start with little communicative ability: the initial \trait{directional bias} is
drawn from $\mathcal{U}(0,0.15)$, \trait{receiver attention} from $\mathcal{U}(0,0.25)$,
the \trait{foray distance} from $\mathcal{U}(0.15D, 0.45D)$, the \trait{dance propensity} from
$\mathcal{U}(0.8,1.0)$, and both \trait{transposition} traits start at $0$, so early
foragers signal weakly, rarely attend to one another, and use only direct
pointing.

\begin{table}
  \centering
  \small
  \begin{tabular}{@{}lr p{0.45\textwidth}@{}}
    \toprule
    Symbol / name & Value & Meaning \\
    \midrule
    colonies & $60$ & population size \\
    workers per colony & $80$ & ensemble size per colony \\
    episodes per colony & $50$ & foraging episodes averaged for fitness \\
    attempts per episode & $12$ & foraging attempts per episode \\
    $\eta$ & $0.08$ & worker-variation scale (trait sd; traits on $[0,1]$) \\
    $\kappa_{\max}$ & $14$ & maximum dance concentration ($\approx 16^\circ$ scatter at $b_i = 1$) \\
    dance noise & $0.18$ & production noise (wrapped Gaussian, rad; $\approx 10^\circ$) \\
    interpretation noise & $0.12$ & receiver decoding noise (rad; $\approx 7^\circ$) \\
    $D$ & $6$ & maximum search distance (km) \\
    $d_{\min}$ & $0.75$ & minimum food distance (km) \\
    $\sigma_r$ & $0.6$ & patch-radius log-scale spread ($80\%$ of radii in $70$--$325$ m) \\
    foray shape & $2$ & Gamma shape of per-foray distance (coefficient of variation $\approx 0.71$) \\
    $c_{\mathrm{base}},\,c_{\mathrm{cue}}$ & $0,\ 0.02$ & dance cost terms (in units of $v$) \\
    $c_{\mathrm{att}}$ & $0.01$ & attention cost (in units of $v$) \\
    $v$ & $1$ & food value (sets the payoff scale) \\
    $\mu_{\mathrm{sun}},\,\Delta_{\mathrm{sun}}$ & $\pi/2,\ \pi$ & sun arc center and width (rad; a $180^\circ$ arc) \\
    \bottomrule
  \end{tabular}
  \caption{Parameters held fixed across all experiments. The parameters that some
    experiment varies are listed separately in \Cref{tab:varied-params}.}
  \label{tab:fixed-params}
\end{table}

The parameters that are not fixed in this way are summarized in
\Cref{tab:varied-params}: five ecological parameters (food-site count, patch
radius, capacity, maximum food distance, and travel cost) and four evolutionary
parameters (the number of generations, $B$, $\sigma_m$, and $\rho$). That table
gives only what each parameter means: the values they take, swept or held fixed,
are stated with the experiments themselves in
\Cref{sec:setup-horizontal,sec:setup-transition}.

\begin{table}
  \centering
  \begin{tabular}{ll}
    \toprule
    Symbol / name & Meaning \\
    \midrule
    \multicolumn{2}{l}{\emph{Ecological}} \\
    food-site count & Poisson mean number of patches per episode \\
    $\mu_r$        & median patch radius (m, lognormal) \\
    capacity        & base loads a patch supplies at the median radius \\
    $d_{\max}$      & maximum patch distance (km) \\
    travel cost     & cost per km of foray distance (in units of $v$) \\
    \midrule
    \multicolumn{2}{l}{\emph{Evolutionary}} \\
    generations & evolutionary horizon \\
    $B$        & vertical-comb benefit magnitude \\
    $\sigma_m$ & mutation step standard deviation \\
    $\rho$     & sender--receiver mutation correlation \\
    \bottomrule
  \end{tabular}
  \caption{Parameters that are not held fixed across all experiments, and what
    each means. No single experiment varies all of them;
    \Cref{sec:setup-horizontal,sec:setup-transition} give the values each takes
    in each experiment, whether swept or held fixed.}
  \label{tab:varied-params}
\end{table}

\subsection{Emergence of direct pointing}
\label{sec:setup-horizontal}
The first stage uses the horizontal-comb model of \Cref{sec:horizontal}: comb
tilt is fixed at $0$ and not allowed to evolve, so the gravity reference has zero
strength and the \trait{transposition} traits are inert, leaving only the direct-pointing
code in play. Colonies evolve for $60$ generations under mutation scale
$\sigma_m = 0.07$, and we vary only the food distribution across $50$ replicate
seeds ($400$--$449$). The conditions form a full grid crossing the Poisson mean
patch count $\{1,2,3,4,6,8,12,16,24\}$ against the median patch radius
$\{15, 37.5, 75, 150, 300, 600\}$ meters, so that the ecology ranges from single small
patches (essentially undiscoverable) to many large ones (easily found by random
search). The remaining parameters of \Cref{tab:varied-params} are held fixed: a
patch at the median radius has capacity $6$, and patches lie no farther than
$d_{\max} = 6$ km. Each km flown subtracts a fixed amount from the colony's
payoff, whether or not the foray finds food. Reaching the nearest patch costs
$2\%$ of the food it yields; reaching the farthest costs $16\%$.

The primary outcome is the \emph{recruitment advantage} (\Cref{sec:foraging}).
Because whether a worker follows a dance is a probabilistic coin flip on
\trait{receiver attention}, comparing followers and searchers within the same
episode approximates a randomized trial of how much benefit the dance
provides, distinguishing functioning communication from a directional-bias
trait that has merely drifted upward. We report it as a mean over
the final generations, alongside final-generation mean \trait{directional bias}, foraging
success, and \trait{dance propensity}.

\subsection{Evolution of the gravity code}
\label{sec:setup-transition}
The second stage uses the full model of \Cref{sec:vertical}: \trait{comb tilt} $\gamma$
and orientation $\phi$ are free to evolve, the \trait{transposition} traits $t_s, t_r$ are
active, and the raw payoff is scaled by the vertical-comb benefit
$1 + B\gamma$. Populations always start horizontal, so a successful run must
evolve both a vertical comb \emph{and} a matched sender--receiver gravity code.
The transition is governed by the interaction of the ecology with three
evolutionary parameters: the benefit magnitude $B$, the mutation scale
$\sigma_m$, and the sender--receiver coupling $\rho$. 

The question we ask in this stage is whether there is \emph{any} combination of
ecology and evolutionary regime which lets a population cross between the two
codes without a breakdown of communication, and if so which. The space in which
to look is both large and costly to cover: eight parameters interact, the values
we admit for them (\Cref{tab:search-space}) already combine into some $2.3$
billion distinct settings, and pricing a single setting costs a full evolutionary
run on a panel of seeds. We therefore use optimization, which spends its budget
where success looks likely instead of spreading it evenly.
The search tells us \emph{where} viable settings lie, but its
trials cluster around what already worked, so their density cannot tell us how
\emph{large} a viable region is. The experiments below therefore re-examine
what the search finds on fixed grids. They also re-examine it on fresh random
seeds, because a search that maximizes over a panel of seeds will fit the
noise of that panel along with its signal.

\paragraph{Decode variants.}
On a tilted comb (\Cref{sec:vertical}) the direct code can be decoded via
\emph{flatten} or \emph{unproject}, and we run the experiments below with each
variant.  The two pipelines are identical in every other respect, so comparing
them isolates the effect of the geometric decode assumption on whether and where
the transition occurs. Which way that effect should run is not clear a priori.
On the one hand, \emph{flatten} is biased on a tilted comb, and this imprecision
may push a colony toward the gravity referenced code. On the other hand,
\emph{unproject} keeps direct pointing accurate as the comb tilts and so keeps
transitional colonies more viable, which may instead ease the transition by
keeping them alive long enough to complete it. 

\paragraph{Parameter search.}
We search eight parameters: food-site count, patch radius, capacity, maximum
distance, and travel cost, together with $B$, $\sigma_m$, and $\rho$.
\Cref{tab:search-space} gives the interval and step size searched for each, and
the sampler proposes values on those steps, so the settings form a
grid rather than a continuum. One \emph{trial} is one such setting, scored by
simulating it as described below; colonies evolve for $120$ generations
throughout. Trials are proposed by the Tree-structured Parzen Estimator
\citep{bergstra2011algorithms} as implemented in Optuna
\citep{akiba2019optuna}, which fits per-parameter distributions to the
best-scoring trials so far and proposes settings typical of them, so that later
trials concentrate where earlier ones succeeded; the first $64$ trials are drawn
at random. We run $1024$ trials per decode variant.

\paragraph{Scoring a trial.}
Each trial evaluates a panel of $J = 10$ seeds. For seed $j$ we
write $\bar\gamma_j$, $\bar t_{s,j}$ and $\bar t_{r,j}$ for the final-generation
population means of the \trait{comb tilt} and the two \trait{transposition}
traits, and $m_j$ for the lowest foraging success (\Cref{sec:foraging}, averaged
over the colonies) reached at any generation of the run. A seed is \emph{stable} if $\bar\gamma_j \ge
\gamma^\ast$ and $\min(\bar t_{s,j}, \bar t_{r,j}) \ge t^\ast$, and
\emph{non-viable} if $m_j \le m^\ast$. The first two thresholds mark what we count
as a completed transition: $\gamma^\ast = 0.80$ asks for a mostly vertical comb,
and $t^\ast = 0.50$ asks that both roles weight the gravity reference at least as
heavily as direct pointing, so that the code has genuinely tipped over rather than
merely drifted. The threshold $m^\ast = 0.02$ is a viability floor:
near-random search leaves the founding generation with low success, but the
minimum can also occur later, when comb tilt outpaces the \trait{transposition}
traits and degrades the ancestral code before the gravity-referenced one takes
over. Penalizing $m_j$ wherever it falls steers the search away from settings in
which foraging is never adequate. The \emph{progress} of a seed measures how near it
came to the first two: each trait is expressed as a fraction of the threshold it must reach,
and progress is the smallest of those fractions, capped at one,
\begin{equation}
  g_j = \min\Bigl(1,\; \frac{\bar\gamma_j}{\gamma^\ast},\;
    \frac{\bar t_{s,j}}{t^\ast},\; \frac{\bar t_{r,j}}{t^\ast}\Bigr),
\end{equation}
so that $g_j$ tracks whichever of the three lags furthest behind. Each fraction
reaches one exactly when its trait reaches its threshold, so $g_j = 1$ holds
exactly when seed $j$ is stable, and $g_j < 1$ grades the near misses. The trial
score is then
\begin{equation}
  \underbrace{\sum_{j=1}^{J} \mathbf{1}\bigl[g_j = 1\bigr]}_{\text{stable count}}
  \;+\; \underbrace{\frac{1}{J}\sum_{j=1}^{J} g_j}_{\text{mean progress}}
  \;-\; \underbrace{\sum_{j=1}^{J} \mathbf{1}\bigl[m_j \le m^\ast\bigr]}_{\text{non-viable count}}.
\end{equation}
Because the progress term is an average of quantities bounded by
one, it can never contribute more than a single point, so one further stable seed
outweighs any improvement in progress across the whole panel: the search pursues
transitions that complete, and uses progress only to rank panels that fail.

\begin{table}
  \centering
  \begin{tabular}{lrr}
    \toprule
    Symbol / name & Interval & Step \\
    \midrule
    food-site count & $1$--$8$ & integer \\
    $\mu_r$ (m)    & $60$--$360$ & $15$ \\
    capacity        & $2$--$12$ & integer \\
    $d_{\max}$ (km) & $3$--$7.5$ & $0.25$ \\
    travel cost (per km) & $0.010$--$0.060$ & $0.0025$ \\
    $B$        & $0.10$--$0.60$ & $0.02$ \\
    $\sigma_m$ & $0.04$--$0.14$ & $0.01$ \\
    $\rho$     & $0.0$--$1.0$   & $0.1$ \\
    \bottomrule
  \end{tabular}
  \caption{The space searched by the optimization. Each parameter is proposed on
    a lattice of the given step within its interval, so the space is discrete
    rather than continuous.}
  \label{tab:search-space}
\end{table}

\paragraph{Confirmation and held-out validation.}
Because the optimization ranks trials by their score on only ten seeds, a top-ranked
candidate may owe that score to a genuinely robust transition or to a lucky draw
of those seeds, and re-scoring it on unseen seeds tells the two
apart. The highest-scoring search trials are re-run on a disjoint confirmation panel of
$40$ seeds, and the best candidates are then validated on $100$
held-out seeds that were used in neither search nor confirmation. Validation reports, per
candidate, the fraction of stable seeds, the non-viable count, and
final-generation means of foraging success, \trait{comb tilt}, and the lower of
the two \trait{transposition} traits.

\paragraph{One-parameter sensitivity.}
The search locates a viable setting but, its trials clustering around what
already worked, does not say how wide the viable region around it is. Around the
strongest validated candidate we therefore perturb each parameter in turn,
holding the others at their validated values, over the same $100$ held-out seeds;
sweeping one axis at a time on a regular grid measures how far
each parameter can move before the transition degrades, and so which parameters
it tolerates freely and which it depends on finely.

\paragraph{Evolutionary-parameter interaction.}
To ask whether mutation parameters can compensate for a weaker architectural
benefit, we hold the validated ecology fixed and run a full-factorial grid over
the three evolutionary parameters, sweeping the benefit across its full search
range while centring the mutation-parameter grids on the values found stable
above,
$B \in \{0.10, 0.30, 0.45, 0.60\}$, $\sigma_m \in \{0.05, 0.07, 0.09, 0.11\}$, and
$\rho \in \{0.0, 0.3, 0.6, 0.9\}$, evaluating each of the $64$ cells over the same
$100$ held-out seeds.

\paragraph{Outcome measures.}
Across the transition experiments we summarize runs by the same thresholds used in
the search objective: the \emph{stable} fraction (vertical comb with a coordinated
gravity code) and the \emph{non-viable} fraction (foraging success $\le 0.02$ at its
worst generation), alongside
final-generation trait and success means. Seed-bootstrap intervals over the
per-seed outcomes quantify sampling uncertainty in the reported fractions.

\section{Results}
\label{sec:results}

We report the two stages in turn: the horizontal-comb ecology which fixes when the
direct-pointing code is worth maintaining (\Cref{sec:setup-horizontal}), and the
vertical transition under both decode variants (\Cref{sec:setup-transition}).

\subsection{Emergence of direct pointing}
\label{sec:results-horizontal}
\begin{figure}
  \centering
  \includegraphics[width=0.9\textwidth]{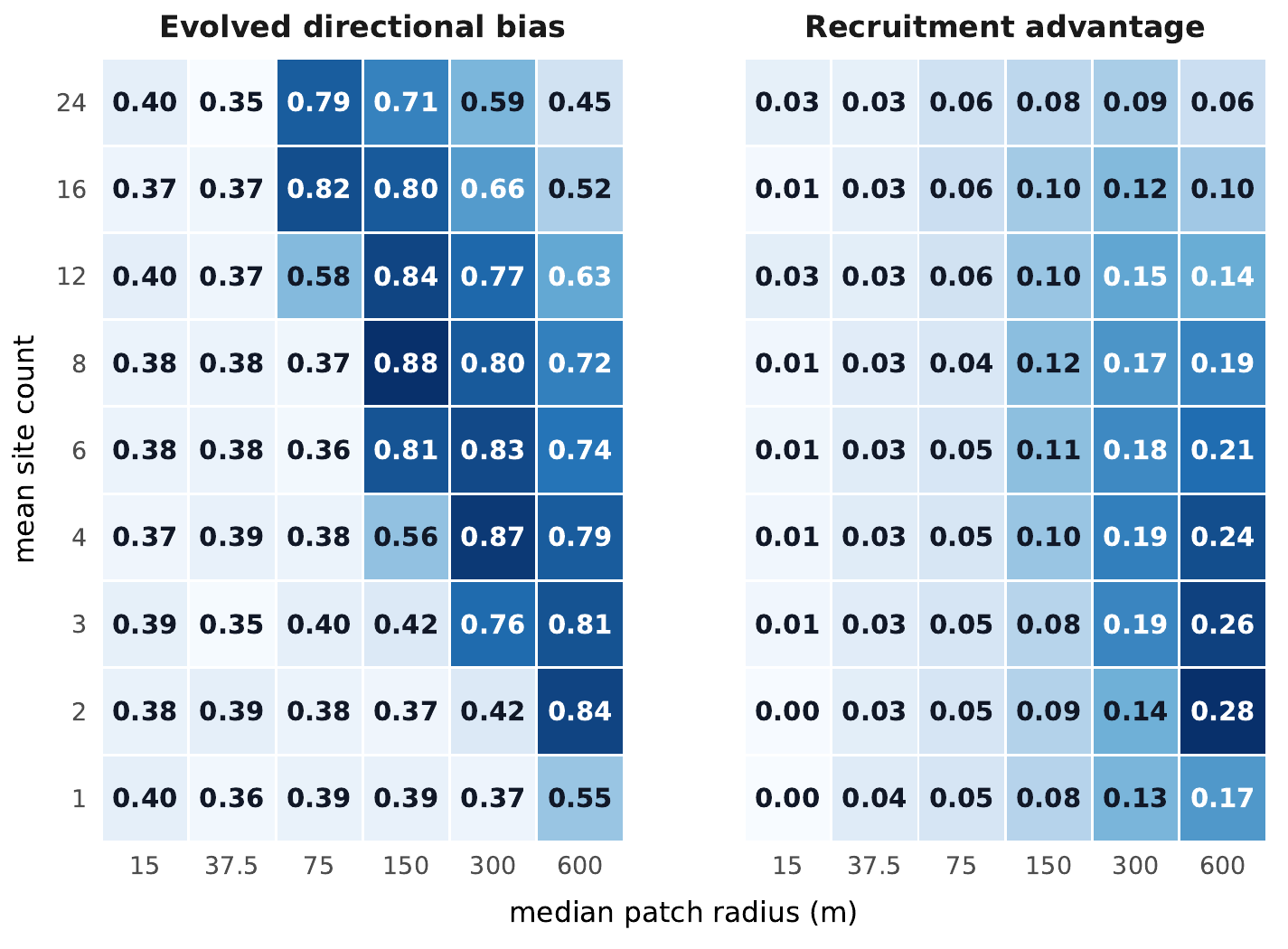}
  \caption{Evolved \trait{directional bias} (left) and recruitment advantage (right) on a
    flat comb across the grid of Poisson mean site count against median patch
    radius, $50$ seeds per cell. Recruitment advantage is the follower--searcher
    success difference, and shading is normalized within each panel.}
  \label{fig:food-grid}
\end{figure}

On a comb held flat, the evolved \trait{directional bias}, which sets how precisely
a dance points, is what tells us communication is favored. Absent selection for
communication it drifts to a low baseline; it rises clearly above that baseline
only along a diagonal band of the food grid (\Cref{fig:food-grid}), and the
patch size at which a precise dance becomes worthwhile falls as patches grow more
numerous. Below a patch radius of a few tens of meters the dance never evolves,
however abundant the food. The recruitment advantage shows what a dance is worth
where it does evolve: it is largest for few, large patches and erodes toward both
small patches, where successful foragers are too rare to seed dances, and abundant
large ones, where independent search already finds the food.

\subsection{Evolution of the gravity code}
\label{sec:results-transition}

\paragraph{Parameter search.}
Stable vertical-gravity transitions, in which a seed ends with a mostly vertical
comb and both worker roles committed to the gravity reference
(\Cref{sec:setup-transition}), are common under both decodes
(\Cref{tab:optuna-stability}). Both searches
concentrate their strongly stable trials, those stable in at least eight of their
ten seeds, in much the same region: abundant, large, reachable food paired with a
strong tilt incentive and a high mutation scale (\Cref{tab:optuna-region}). The
main difference is that \emph{flatten}'s strongly stable trials occur at the edge
of the search range on patch radius and travel cost, needing larger patches and
cheaper travel than \emph{unproject} to remain viable.

\begin{table}
  \centering
  \begin{tabular}{lrr}
    \toprule
    Outcome & Flatten trials & Unproject trials \\
    \midrule
    Stable in all $10$ seeds & $61$ & $33$ \\
    Stable in $\ge 8$ seeds   & $252$ & $214$ \\
    Stable in $\ge 5$ seeds   & $369$ & $375$ \\
    Stable in $\ge 1$ seed    & $655$ & $674$ \\
    No stable seed            & $369$ & $350$ \\
    \bottomrule
  \end{tabular}
  \caption{Stability counts over the $1024$ Optuna trials (ten seeds each) under
    each decode.}
  \label{tab:optuna-stability}
\end{table}

\begin{table}
  \centering
  \small
  \begin{tabular}{lrlrl}
    \toprule
    & \multicolumn{2}{c}{Flatten} & \multicolumn{2}{c}{Unproject} \\
    \cmidrule(lr){2-3}\cmidrule(lr){4-5}
    Parameter & Median & 10th--90th & Median & 10th--90th \\
    \midrule
    Food-site count (Poisson mean) & $7$    & $4$--$8$   & $7$    & $4$--$8$ \\
    Patch radius (m)               & $345$  & $255$--$360$ & $285$  & $215$--$360$ \\
    Patch capacity                 & $4$    & $2$--$9$   & $4$    & $2$--$9$ \\
    Vertical-comb benefit $B$      & $0.58$ & $0.54$--$0.60$ & $0.54$ & $0.50$--$0.60$ \\
    Max food distance (km)         & $3.75$ & $3.25$--$5.25$ & $3.75$ & $3.00$--$5.25$ \\
    Travel cost per km             & $0.013$ & $0.010$--$0.017$ & $0.020$ & $0.010$--$0.045$ \\
    Mutation scale $\sigma_m$      & $0.100$ & $0.090$--$0.110$ & $0.110$ & $0.090$--$0.130$ \\
    Correlation $\rho$             & $0.9$   & $0.0$--$1.0$ & $0.6$   & $0.0$--$1.0$ \\
    \bottomrule
  \end{tabular}
  \caption{Parameter distribution of the strongly stable trials (eight or more of
    ten stable seeds) under each decode.}
  \label{tab:optuna-region}
\end{table}

\paragraph{Held-out validation.}
The top five distinct candidates emerging from the $40$-seed confirmation
panel of each search, rerun on $100$ held-out seeds
($200$--$299$), all produced frequent stable transitions and no non-viable seeds,
with the flatten candidates achieving somewhat higher stable rates than the
unproject ones. The strongest candidate of each decode (\Cref{tab:validation}) lands
on similar held-out rates and on overlapping ecologies: large patches,
short-to-moderate distances, low travel cost, high benefit.

\begin{table}
  \centering
  \begin{tabular}{lrr}
    \toprule
    Parameter & Flatten & Unproject \\
              & (cand.\ 626) & (cand.\ 977) \\
    \midrule
    Food-site count (Poisson mean) & $5$    & $8$ \\
    Patch radius (m)               & $345$  & $345$ \\
    Patch capacity                 & $9$    & $2$ \\
    Vertical-comb benefit $B$      & $0.60$ & $0.56$ \\
    Max food distance (km)         & $5.25$ & $4.25$ \\
    Travel cost per km             & $0.018$ & $0.020$ \\
    Mutation scale $\sigma_m$      & $0.100$ & $0.090$ \\
    Correlation $\rho$             & $1.0$   & $0.4$ \\
    \addlinespace
    Stable seeds (of $100$)        & $95$    & $90$ \\
    Final mean \trait{comb tilt} $t_f$ & $0.848$ & $0.846$ \\
    \bottomrule
  \end{tabular}
  \caption{Strongest validated candidate per decode over $100$ held-out seeds.}
  \label{tab:validation}
\end{table}

\begin{figure}
  \centering
  \includegraphics[width=0.82\textwidth]{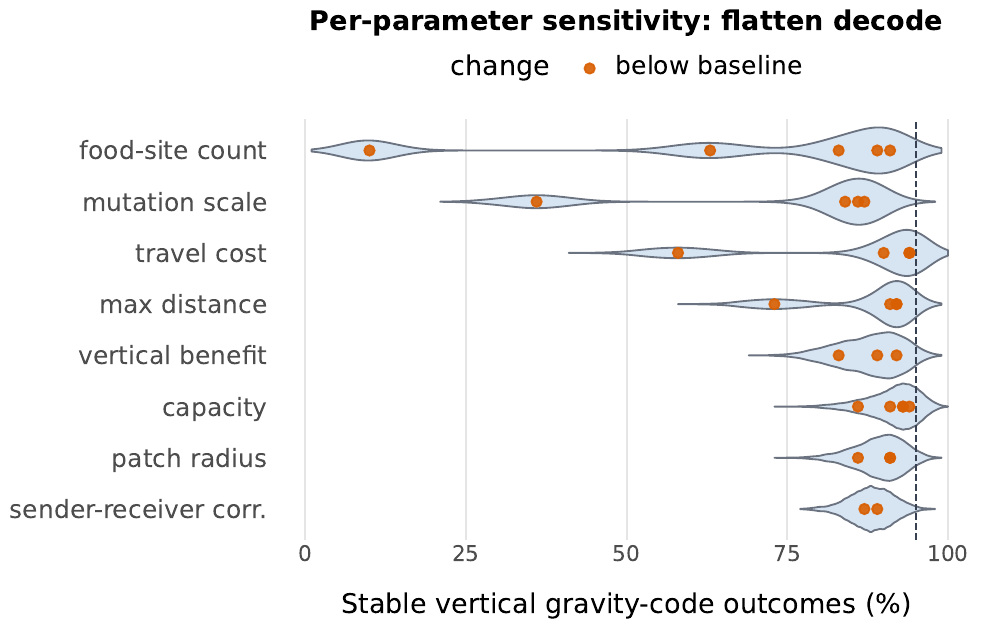}
  \caption{One-parameter sensitivity of the flatten decode. The violin for each parameter
    is a kernel density over seed-bootstrap stable-rate draws pooled across the
    swept values of that parameter ($100$ held-out seeds per value), so each lobe
    is centered on the rate for one swept value, with its width showing the
    sampling uncertainty of that value; dots mark the per-value rates (orange
    below, green above baseline) and the dashed line the baseline. Parameters are sorted by worst-case drop (shared order and x-range
    with \Cref{fig:sensitivity-unproject}).}
  \label{fig:sensitivity-flatten}
\end{figure}

\begin{figure}
  \centering
  \includegraphics[width=0.82\textwidth]{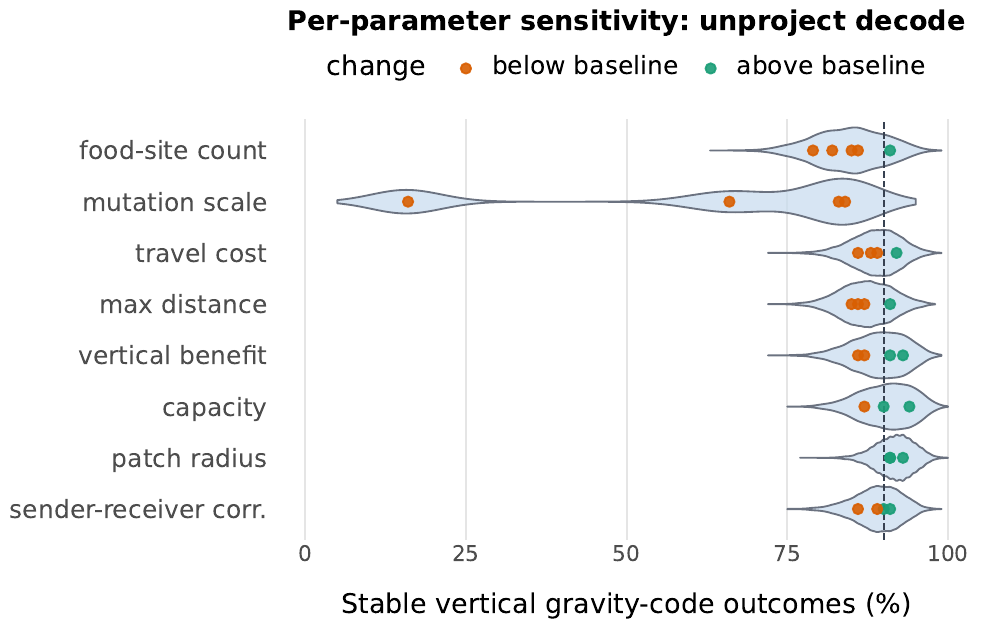}
  \caption{One-parameter sensitivity of the unproject decode, drawn as in
    \Cref{fig:sensitivity-flatten} and sharing its parameter order and x-range.}
  \label{fig:sensitivity-unproject}
\end{figure}

\paragraph{One-parameter sensitivity.}
A one-parameter sweep around each validated baseline finds a different sharpest
boundary per decode (\Cref{fig:sensitivity-flatten,fig:sensitivity-unproject}).
Under unproject, mutation scale is the one parameter whose perturbation strongly
affects the outcome; every other parameter, food-site count included, makes
little difference. Under flatten, sensitivity is spread across several
parameters: food-site count matters most, followed by mutation scale, then
travel cost and maximum distance. Meanwhile patch radius, capacity,
vertical-comb benefit, and correlation make little difference. Most food lies at
long range, where the tilt-dependent decoding bias of flatten makes precise
dances hard to produce, pushing its validated point to the edge of the search range on
patch radius and travel cost, with little margin left elsewhere.

\begin{figure}
  \centering
  \includegraphics[width=\textwidth]{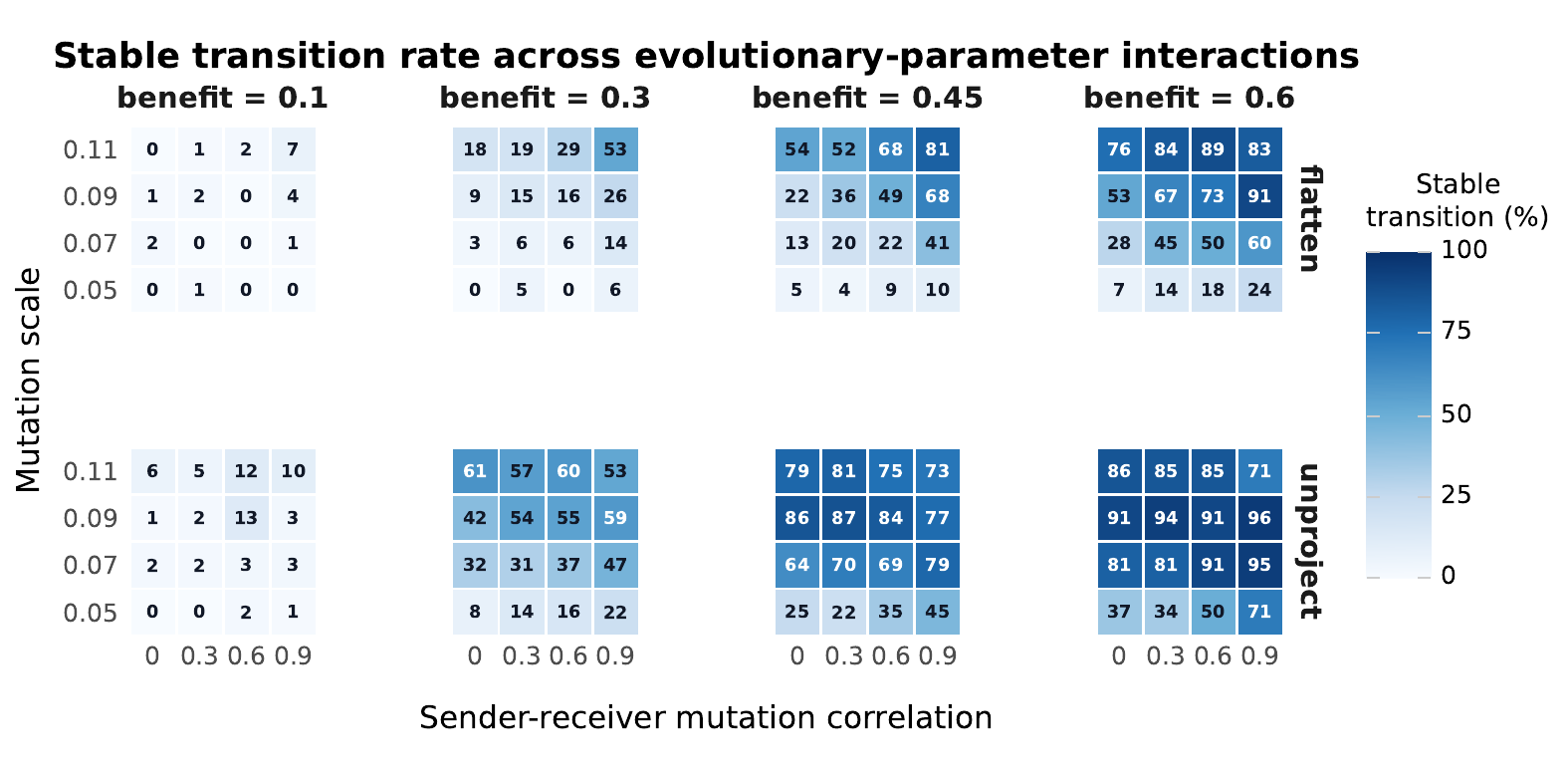}
  \caption{Stable transition rate across the evolutionary-parameter interaction
    grid, holding the validated ecology of each decode fixed. Benefit is swept across
    the full search range ($0.10$--$0.60$) as a generic gradient rather than
    through the validated operating points (flatten $B{=}0.56$, unproject
    $B{=}0.60$). Tiles cross the sender--receiver mutation correlation $\rho$
    against the mutation scale $\sigma_m$, faceted by decode (rows) and
    vertical-comb benefit $B$ (columns); each cell summarizes $100$ held-out
    seeds, and its label is the percentage of seeds with a stable transition.}
  \label{fig:interaction}
\end{figure}

\paragraph{Evolutionary-parameter interaction.}
Fixing the validated ecology of each decode and sweeping $B$, $\sigma_m$, and $\rho$
across the search range, stable rate rises with benefit for both decodes
(\Cref{fig:interaction}). A complete failure to produce even one stable
transition among the $100$ seeds is rare for both decodes, but noticeably rarer
and more constrained for unproject. Within each benefit panel, higher mutation
scale and correlation both help, but only partly compensate for weak benefit.

\section{Discussion and Conclusion}
\label{sec:discussion}

The current study poses two questions: under what ecological conditions a
direct-pointing referential code emerges and stabilizes, and which factors let
a population move from this code to its gravity-referenced variant. Both are
feasibility questions rather than probability estimates: they ask what
conditions make each code possible at all, not how likely a given ecology or
lineage was to realize it. The two stages of the simulation answer them in
turn.

On a horizontal comb, direct pointing is feasible only within a moderate band
of foraging difficulty (\Cref{fig:food-grid}), and the edges of that band are
gradual rather than sharp: too little food never seeds enough dances to be
useful, too much makes independent search sufficient on its own. The latter effect
mirrors the findings from foraging simulations
\citep{dornhaus2006benefits,beekman2008pay}, but now in the context of
evolving populations of communicating agents.

The transition itself turns out to be feasible quite broadly. Sweeping the
benefit $B$, the mutation scale $\sigma_m$, and the sender--receiver
correlation $\rho$ over their full search ranges at a fixed, favorable
ecology, a transition fails completely only in a small minority of grid
cells, concentrated at low benefit. In a large majority of cells, transition
occurs in some seeds, though often only in a minority. The regime the search
converges on, a high mutation scale paired with a large benefit, is not
especially biologically plausible, nor is it necessary to allow the code
transition to occur.

We interpret this pattern of results as a demonstration of the feasibility of
transition, rather than a claim about its likelihood: across nearly all of
the range we tested, nothing about the evolutionary parameters rules the
transition out.

Taken together, these answers complement the mechanistic proposal of
\citet{10.1242/jeb.142778}, which addresses how a single bee's brain can
support either code: if the underlying heading representation is indeed
indifferent to the spatial reference frame, our results indicate where the
remaining barrier lies, not in individual cognition but in whether an
evolving population of senders and receivers can stay coordinated as the code
itself changes.

The broader picture across both stages is the same kind of claim at two
different scales: a referential code, and later its replacement, need not be
finely tuned into existence. Direct pointing is favored across a real, if
bounded, swath of foraging ecologies; the shift to a gravity-referenced code,
once that ecology holds and vertical building pays off, fails only at low
benefit, remaining open everywhere else we tested. This is a claim about what is
possible, not what was likely in the actual evolutionary history of \emph{Apis}:
that would need independent evidence on ancestral foraging ecology, the real
benefit of vertical combs, and the genetic basis of dance production and
interpretation. Feasibility is, however, enough to answer the coordination
problem posed in the introduction: a shared code can be revised without a
breakdown of communication across most of the conditions we examined, not just
at one finely tuned point among them.

\section{Limitations}
\label{sec:limitations}

Several aspects of the model bound what these results can support. Only the
directional component of the dance is modeled, so the results say nothing about how
encoding distance would affect the emergence or the transition of the code.
Patches differ in size and distance, but not in quality, so a dance can send
a follower to a patch, but never to a better one. The recruitment advantage we
report therefore omits what other work identifies as an important benefit of the
dance: concentrating foragers on the most profitable sites
\citep{beekman2008pay,donaldsonmatasci2012habitat}.

Colonies reproduce asexually, with no gene flow between them. The
vertical-comb benefit $B$ is an exogenous, fixed-form term
(\Cref{sec:transition}) rather than one derived from a mechanical or thermal
account of comb architecture, so the model is silent on the magnitude or
origin of the real advantage. Likewise, the mutation scale $\sigma_m$ sets how fast
colony traits explore trait space over a fixed number of generations, not a
real per-generation mutation rate. Transposition is also blended continuously rather than
switched discretely (\Cref{sec:vertical}): a permissive choice that may widen
the feasible region we report.

The parameter analyses are similarly bounded. The one-parameter sensitivity
sweeps (\Cref{fig:sensitivity-flatten,fig:sensitivity-unproject}) and the
evolutionary-parameter interaction grid (\Cref{fig:interaction}) both hold
ecology fixed at a single validated candidate, so neither shows how these
effects generalize elsewhere; and because the horizontal and transition stages
sweep different ecological grids under different evolutionary horizons
(\Cref{sec:setup-horizontal,sec:setup-transition}), their results are best read
side by side rather than combined into a single claim.

\bibliographystyle{apacite}
\bibliography{biblio}

\begin{thebibliography}{}

\bibitem [\protect \citeauthoryear {%
Akiba%
, Sano%
, Yanase%
, Ohta%
\BCBL {}\ \BBA {} Koyama%
}{%
Akiba%
\ \protect \BOthers {.}}{%
{\protect \APACyear {2019}}%
}]{%
akiba2019optuna}
\APACinsertmetastar {%
akiba2019optuna}%
\begin{APACrefauthors}%
Akiba, T.%
, Sano, S.%
, Yanase, T.%
, Ohta, T.%
\BCBL {}\ \BBA {} Koyama, M.%
\end{APACrefauthors}%
\unskip\
\newblock
\APACrefYearMonthDay{2019}{}{}.
\newblock
{\BBOQ}\APACrefatitle {Optuna: A Next-generation Hyperparameter Optimization Framework} {Optuna: A next-generation hyperparameter optimization framework}.{\BBCQ}
\newblock
\BIn{} \APACrefbtitle {Proceedings of the 25th ACM SIGKDD International Conference on Knowledge Discovery \& Data Mining} {Proceedings of the 25th acm sigkdd international conference on knowledge discovery \& data mining}\ (\BPGS\ 2623--2631).
\newblock
\APACaddressPublisher{}{ACM}.
\newblock
\begin{APACrefDOI} \doi{10.1145/3292500.3330701} \end{APACrefDOI}
\PrintBackRefs{\CurrentBib}

\bibitem [\protect \citeauthoryear {%
Bailis%
, Nagpal%
\BCBL {}\ \BBA {} Werfel%
}{%
Bailis%
\ \protect \BOthers {.}}{%
{\protect \APACyear {2010}}%
}]{%
bailis2010positional}
\APACinsertmetastar {%
bailis2010positional}%
\begin{APACrefauthors}%
Bailis, P.%
, Nagpal, R.%
\BCBL {}\ \BBA {} Werfel, J.%
\end{APACrefauthors}%
\unskip\
\newblock
\APACrefYearMonthDay{2010}{}{}.
\newblock
{\BBOQ}\APACrefatitle {Positional Communication and Private Information in Honeybee Foraging Models} {Positional communication and private information in honeybee foraging models}.{\BBCQ}
\newblock
\BIn{} \APACrefbtitle {Swarm Intelligence: 7th International Conference, ANTS 2010} {Swarm intelligence: 7th international conference, ants 2010}\ (\BVOL\ 6234, \BPGS\ 263--274).
\newblock
\APACaddressPublisher{}{Springer}.
\newblock
\begin{APACrefDOI} \doi{10.1007/978-3-642-15461-4_23} \end{APACrefDOI}
\PrintBackRefs{\CurrentBib}

\bibitem [\protect \citeauthoryear {%
Barron%
\ \BBA {} Plath%
}{%
Barron%
\ \BBA {} Plath%
}{%
{\protect \APACyear {2017}}%
}]{%
10.1242/jeb.142778}
\APACinsertmetastar {%
10.1242/jeb.142778}%
\begin{APACrefauthors}%
Barron, A\BPBI B.%
\BCBT {}\ \BBA {} Plath, J\BPBI A.%
\end{APACrefauthors}%
\unskip\
\newblock
\APACrefYearMonthDay{2017}{12}{}.
\newblock
{\BBOQ}\APACrefatitle {The evolution of honey bee dance communication: a mechanistic perspective} {The evolution of honey bee dance communication: a mechanistic perspective}.{\BBCQ}
\newblock
\APACjournalVolNumPages{Journal of Experimental Biology}{220}{23}{4339-4346}.
\newblock
\begin{APACrefURL} \url{https://doi.org/10.1242/jeb.142778} \end{APACrefURL}
\newblock
\begin{APACrefDOI} \doi{10.1242/jeb.142778} \end{APACrefDOI}
\PrintBackRefs{\CurrentBib}

\bibitem [\protect \citeauthoryear {%
Beekman%
\ \BBA {} Lew%
}{%
Beekman%
\ \BBA {} Lew%
}{%
{\protect \APACyear {2008}}%
}]{%
beekman2008pay}
\APACinsertmetastar {%
beekman2008pay}%
\begin{APACrefauthors}%
Beekman, M.%
\BCBT {}\ \BBA {} Lew, J\BPBI B.%
\end{APACrefauthors}%
\unskip\
\newblock
\APACrefYearMonthDay{2008}{}{}.
\newblock
{\BBOQ}\APACrefatitle {Foraging in honeybees---when does it pay to dance?} {Foraging in honeybees---when does it pay to dance?}{\BBCQ}
\newblock
\APACjournalVolNumPages{Behavioral Ecology}{19}{2}{255--262}.
\newblock
\begin{APACrefDOI} \doi{10.1093/beheco/arm117} \end{APACrefDOI}
\PrintBackRefs{\CurrentBib}

\bibitem [\protect \citeauthoryear {%
Beekman%
\ \BBA {} Ratnieks%
}{%
Beekman%
\ \BBA {} Ratnieks%
}{%
{\protect \APACyear {2000}}%
}]{%
beekman2000longrange}
\APACinsertmetastar {%
beekman2000longrange}%
\begin{APACrefauthors}%
Beekman, M.%
\BCBT {}\ \BBA {} Ratnieks, F\BPBI L\BPBI W.%
\end{APACrefauthors}%
\unskip\
\newblock
\APACrefYearMonthDay{2000}{}{}.
\newblock
{\BBOQ}\APACrefatitle {Long-range foraging by the honey-bee, {Apis mellifera} L.} {Long-range foraging by the honey-bee, {Apis mellifera} l.}{\BBCQ}
\newblock
\APACjournalVolNumPages{Functional Ecology}{14}{4}{490--496}.
\newblock
\begin{APACrefDOI} \doi{10.1046/j.1365-2435.2000.00443.x} \end{APACrefDOI}
\PrintBackRefs{\CurrentBib}

\bibitem [\protect \citeauthoryear {%
Bergstra%
, Bardenet%
, Bengio%
\BCBL {}\ \BBA {} K{\'e}gl%
}{%
Bergstra%
\ \protect \BOthers {.}}{%
{\protect \APACyear {2011}}%
}]{%
bergstra2011algorithms}
\APACinsertmetastar {%
bergstra2011algorithms}%
\begin{APACrefauthors}%
Bergstra, J.%
, Bardenet, R.%
, Bengio, Y.%
\BCBL {}\ \BBA {} K{\'e}gl, B.%
\end{APACrefauthors}%
\unskip\
\newblock
\APACrefYearMonthDay{2011}{}{}.
\newblock
{\BBOQ}\APACrefatitle {Algorithms for Hyper-Parameter Optimization} {Algorithms for hyper-parameter optimization}.{\BBCQ}
\newblock
\BIn{} \APACrefbtitle {Advances in Neural Information Processing Systems 24 (NIPS 2011)} {Advances in neural information processing systems 24 (nips 2011)}\ (\BPGS\ 2546--2554).
\newblock
\begin{APACrefURL} \url{https://proceedings.neurips.cc/paper/2011/hash/86e8f7ab32cfd12577bc2619bc635690-Abstract.html} \end{APACrefURL}
\PrintBackRefs{\CurrentBib}

\bibitem [\protect \citeauthoryear {%
Bernard%
, Wischmann%
, Floreano%
\BCBL {}\ \BBA {} Keller%
}{%
Bernard%
\ \protect \BOthers {.}}{%
{\protect \APACyear {2023}}%
}]{%
10.1371/journal.pcbi.1010487}
\APACinsertmetastar {%
10.1371/journal.pcbi.1010487}%
\begin{APACrefauthors}%
Bernard, A.%
, Wischmann, S.%
, Floreano, D.%
\BCBL {}\ \BBA {} Keller, L.%
\end{APACrefauthors}%
\unskip\
\newblock
\APACrefYearMonthDay{2023}{03}{}.
\newblock
{\BBOQ}\APACrefatitle {The evolution of behavioral cues and signaling in displaced communication} {The evolution of behavioral cues and signaling in displaced communication}.{\BBCQ}
\newblock
\APACjournalVolNumPages{PLOS Computational Biology}{19}{3}{1-16}.
\newblock
\begin{APACrefURL} \url{https://doi.org/10.1371/journal.pcbi.1010487} \end{APACrefURL}
\newblock
\begin{APACrefDOI} \doi{10.1371/journal.pcbi.1010487} \end{APACrefDOI}
\PrintBackRefs{\CurrentBib}

\bibitem [\protect \citeauthoryear {%
Bradbury%
\ \BBA {} Vehrencamp%
}{%
Bradbury%
\ \BBA {} Vehrencamp%
}{%
{\protect \APACyear {2011}}%
}]{%
bradbury2011principles}
\APACinsertmetastar {%
bradbury2011principles}%
\begin{APACrefauthors}%
Bradbury, J\BPBI W.%
\BCBT {}\ \BBA {} Vehrencamp, S\BPBI L.%
\end{APACrefauthors}%
\unskip\
\newblock
\APACrefYear{2011}.
\newblock
\APACrefbtitle {Principles of Animal Communication} {Principles of animal communication}\ (\PrintOrdinal{2}\ \BEd).
\newblock
\APACaddressPublisher{Sunderland, MA}{Sinauer Associates}.
\PrintBackRefs{\CurrentBib}

\bibitem [\protect \citeauthoryear {%
Chaabouni%
, Kharitonov%
, Bouchacourt%
, Dupoux%
\BCBL {}\ \BBA {} Baroni%
}{%
Chaabouni%
\ \protect \BOthers {.}}{%
{\protect \APACyear {2020}}%
}]{%
chaabouni2020compositionality}
\APACinsertmetastar {%
chaabouni2020compositionality}%
\begin{APACrefauthors}%
Chaabouni, R.%
, Kharitonov, E.%
, Bouchacourt, D.%
, Dupoux, E.%
\BCBL {}\ \BBA {} Baroni, M.%
\end{APACrefauthors}%
\unskip\
\newblock
\APACrefYearMonthDay{2020}{}{}.
\newblock
{\BBOQ}\APACrefatitle {Compositionality and Generalization in Emergent Languages} {Compositionality and generalization in emergent languages}.{\BBCQ}
\newblock
\BIn{} \APACrefbtitle {Proceedings of the 58th Annual Meeting of the Association for Computational Linguistics (ACL)} {Proceedings of the 58th annual meeting of the association for computational linguistics (acl)}\ (\BPGS\ 4427--4442).
\PrintBackRefs{\CurrentBib}

\bibitem [\protect \citeauthoryear {%
Donaldson-Matasci%
\ \BBA {} Dornhaus%
}{%
Donaldson-Matasci%
\ \BBA {} Dornhaus%
}{%
{\protect \APACyear {2012}}%
}]{%
donaldsonmatasci2012habitat}
\APACinsertmetastar {%
donaldsonmatasci2012habitat}%
\begin{APACrefauthors}%
Donaldson-Matasci, M\BPBI C.%
\BCBT {}\ \BBA {} Dornhaus, A.%
\end{APACrefauthors}%
\unskip\
\newblock
\APACrefYearMonthDay{2012}{}{}.
\newblock
{\BBOQ}\APACrefatitle {How habitat affects the benefits of communication in collectively foraging honey bees} {How habitat affects the benefits of communication in collectively foraging honey bees}.{\BBCQ}
\newblock
\APACjournalVolNumPages{Behavioral Ecology and Sociobiology}{66}{4}{583--592}.
\newblock
\begin{APACrefDOI} \doi{10.1007/s00265-011-1306-z} \end{APACrefDOI}
\PrintBackRefs{\CurrentBib}

\bibitem [\protect \citeauthoryear {%
Dornhaus%
, Kl\"ugl%
, Oechslein%
, Puppe%
\BCBL {}\ \BBA {} Chittka%
}{%
Dornhaus%
\ \protect \BOthers {.}}{%
{\protect \APACyear {2006}}%
}]{%
dornhaus2006benefits}
\APACinsertmetastar {%
dornhaus2006benefits}%
\begin{APACrefauthors}%
Dornhaus, A.%
, Kl\"ugl, F.%
, Oechslein, C.%
, Puppe, F.%
\BCBL {}\ \BBA {} Chittka, L.%
\end{APACrefauthors}%
\unskip\
\newblock
\APACrefYearMonthDay{2006}{}{}.
\newblock
{\BBOQ}\APACrefatitle {Benefits of recruitment in honey bees: effects of ecology and colony size in an individual-based model} {Benefits of recruitment in honey bees: effects of ecology and colony size in an individual-based model}.{\BBCQ}
\newblock
\APACjournalVolNumPages{Behavioral Ecology}{17}{3}{336--344}.
\newblock
\begin{APACrefDOI} \doi{10.1093/beheco/arj036} \end{APACrefDOI}
\PrintBackRefs{\CurrentBib}

\bibitem [\protect \citeauthoryear {%
Dyer%
}{%
Dyer%
}{%
{\protect \APACyear {1985}}%
}]{%
dyer1985mechanisms}
\APACinsertmetastar {%
dyer1985mechanisms}%
\begin{APACrefauthors}%
Dyer, F\BPBI C.%
\end{APACrefauthors}%
\unskip\
\newblock
\APACrefYearMonthDay{1985}{}{}.
\newblock
{\BBOQ}\APACrefatitle {Mechanisms of dance orientation in the Asian honey bee {Apis florea} L.} {Mechanisms of dance orientation in the asian honey bee {Apis florea} l.}{\BBCQ}
\newblock
\APACjournalVolNumPages{Journal of Comparative Physiology A}{157}{}{183--198}.
\newblock
\begin{APACrefDOI} \doi{10.1007/BF01350026} \end{APACrefDOI}
\PrintBackRefs{\CurrentBib}

\bibitem [\protect \citeauthoryear {%
Dyer%
}{%
Dyer%
}{%
{\protect \APACyear {2002}}%
}]{%
dyer2002biology}
\APACinsertmetastar {%
dyer2002biology}%
\begin{APACrefauthors}%
Dyer, F\BPBI C.%
\end{APACrefauthors}%
\unskip\
\newblock
\APACrefYearMonthDay{2002}{}{}.
\newblock
{\BBOQ}\APACrefatitle {The Biology of the Dance Language} {The biology of the dance language}.{\BBCQ}
\newblock
\APACjournalVolNumPages{Annual Review of Entomology}{47}{}{917--949}.
\newblock
\begin{APACrefDOI} \doi{10.1146/annurev.ento.47.091201.145306} \end{APACrefDOI}
\PrintBackRefs{\CurrentBib}

\bibitem [\protect \citeauthoryear {%
Edrich%
}{%
Edrich%
}{%
{\protect \APACyear {1977}}%
}]{%
edrich1977interaction}
\APACinsertmetastar {%
edrich1977interaction}%
\begin{APACrefauthors}%
Edrich, W.%
\end{APACrefauthors}%
\unskip\
\newblock
\APACrefYearMonthDay{1977}{}{}.
\newblock
{\BBOQ}\APACrefatitle {Interaction of light and gravity in the orientation of the waggle dance of honey bees} {Interaction of light and gravity in the orientation of the waggle dance of honey bees}.{\BBCQ}
\newblock
\APACjournalVolNumPages{Animal Behaviour}{25}{}{342--363}.
\newblock
\begin{APACrefDOI} \doi{10.1016/0003-3472(77)90010-0} \end{APACrefDOI}
\PrintBackRefs{\CurrentBib}

\bibitem [\protect \citeauthoryear {%
Esch%
}{%
Esch%
}{%
{\protect \APACyear {2012}}%
}]{%
Esch2012}
\APACinsertmetastar {%
Esch2012}%
\begin{APACrefauthors}%
Esch, H.%
\end{APACrefauthors}%
\unskip\
\newblock
\APACrefYearMonthDay{2012}{}{}.
\newblock
{\BBOQ}\APACrefatitle {Foraging Honey Bees: How Foragers Determine and Transmit Information About Feeding Site Locations} {Foraging honey bees: How foragers determine and transmit information about feeding site locations}.{\BBCQ}
\newblock
\BIn{} C\BPBI G.~Galizia, D.~Eisenhardt\BCBL {}\ \BBA {} M.~Giurfa\ (\BEDS), \APACrefbtitle {Honeybee Neurobiology and Behavior: A Tribute to Randolf Menzel} {Honeybee neurobiology and behavior: A tribute to randolf menzel}\ (\BPGS\ 53--64).
\newblock
\APACaddressPublisher{Dordrecht}{Springer Netherlands}.
\newblock
\begin{APACrefURL} \url{https://doi.org/10.1007/978-94-007-2099-2_5} \end{APACrefURL}
\newblock
\begin{APACrefDOI} \doi{10.1007/978-94-007-2099-2_5} \end{APACrefDOI}
\PrintBackRefs{\CurrentBib}

\bibitem [\protect \citeauthoryear {%
Hamilton%
}{%
Hamilton%
}{%
{\protect \APACyear {1964}}%
}]{%
hamilton1964genetical}
\APACinsertmetastar {%
hamilton1964genetical}%
\begin{APACrefauthors}%
Hamilton, W\BPBI D.%
\end{APACrefauthors}%
\unskip\
\newblock
\APACrefYearMonthDay{1964}{}{}.
\newblock
{\BBOQ}\APACrefatitle {The Genetical Evolution of Social Behaviour. {I}} {The genetical evolution of social behaviour. {I}}.{\BBCQ}
\newblock
\APACjournalVolNumPages{Journal of Theoretical Biology}{7}{1}{1--16}.
\PrintBackRefs{\CurrentBib}

\bibitem [\protect \citeauthoryear {%
Hepburn%
, Pirk%
\BCBL {}\ \BBA {} Duangphakdee%
}{%
Hepburn%
\ \protect \BOthers {.}}{%
{\protect \APACyear {2014}}%
}]{%
hepburn2014nests}
\APACinsertmetastar {%
hepburn2014nests}%
\begin{APACrefauthors}%
Hepburn, H\BPBI R.%
, Pirk, C\BPBI W\BPBI W.%
\BCBL {}\ \BBA {} Duangphakdee, O.%
\end{APACrefauthors}%
\unskip\
\newblock
\APACrefYear{2014}.
\newblock
\APACrefbtitle {Honeybee Nests: Composition, Structure, Function} {Honeybee nests: Composition, structure, function}.
\newblock
\APACaddressPublisher{Berlin, Heidelberg}{Springer}.
\newblock
\begin{APACrefDOI} \doi{10.1007/978-3-642-54328-9} \end{APACrefDOI}
\PrintBackRefs{\CurrentBib}

\bibitem [\protect \citeauthoryear {%
Hepburn%
\ \BBA {} Radloff%
}{%
Hepburn%
\ \BBA {} Radloff%
}{%
{\protect \APACyear {2011}}%
}]{%
hepburn2011biogeography}
\APACinsertmetastar {%
hepburn2011biogeography}%
\begin{APACrefauthors}%
Hepburn, H\BPBI R.%
\BCBT {}\ \BBA {} Radloff, S\BPBI E.%
\end{APACrefauthors}%
\unskip\
\newblock
\APACrefYearMonthDay{2011}{}{}.
\newblock
{\BBOQ}\APACrefatitle {Biogeography of the dwarf honeybees, {Apis andreniformis} and {Apis florea}} {Biogeography of the dwarf honeybees, {Apis andreniformis} and {Apis florea}}.{\BBCQ}
\newblock
\APACjournalVolNumPages{Apidologie}{42}{3}{293--300}.
\newblock
\begin{APACrefDOI} \doi{10.1007/s13592-011-0024-x} \end{APACrefDOI}
\PrintBackRefs{\CurrentBib}

\bibitem [\protect \citeauthoryear {%
Hockett%
}{%
Hockett%
}{%
{\protect \APACyear {1960}}%
}]{%
hockett1960origin}
\APACinsertmetastar {%
hockett1960origin}%
\begin{APACrefauthors}%
Hockett, C\BPBI F.%
\end{APACrefauthors}%
\unskip\
\newblock
\APACrefYearMonthDay{1960}{}{}.
\newblock
{\BBOQ}\APACrefatitle {The Origin of Speech} {The origin of speech}.{\BBCQ}
\newblock
\APACjournalVolNumPages{Scientific American}{203}{3}{88--96}.
\newblock
\begin{APACrefDOI} \doi{10.1038/scientificamerican0960-88} \end{APACrefDOI}
\PrintBackRefs{\CurrentBib}

\bibitem [\protect \citeauthoryear {%
Jackson%
\ \BBA {} Ratnieks%
}{%
Jackson%
\ \BBA {} Ratnieks%
}{%
{\protect \APACyear {2006}}%
}]{%
jackson2006communication}
\APACinsertmetastar {%
jackson2006communication}%
\begin{APACrefauthors}%
Jackson, D\BPBI E.%
\BCBT {}\ \BBA {} Ratnieks, F\BPBI L\BPBI W.%
\end{APACrefauthors}%
\unskip\
\newblock
\APACrefYearMonthDay{2006}{}{}.
\newblock
{\BBOQ}\APACrefatitle {Communication in ants} {Communication in ants}.{\BBCQ}
\newblock
\APACjournalVolNumPages{Current Biology}{16}{15}{R570--R574}.
\newblock
\begin{APACrefDOI} \doi{10.1016/j.cub.2006.07.015} \end{APACrefDOI}
\PrintBackRefs{\CurrentBib}

\bibitem [\protect \citeauthoryear {%
Kohl%
\ \protect \BOthers {.}}{%
Kohl%
\ \protect \BOthers {.}}{%
{\protect \APACyear {2020}}%
}]{%
kohl2020dialects}
\APACinsertmetastar {%
kohl2020dialects}%
\begin{APACrefauthors}%
Kohl, P\BPBI L.%
, Thulasi, N.%
, Rutschmann, B.%
, George, E\BPBI A.%
, Steffan-Dewenter, I.%
\BCBL {}\ \BBA {} Brockmann, A.%
\end{APACrefauthors}%
\unskip\
\newblock
\APACrefYearMonthDay{2020}{}{}.
\newblock
{\BBOQ}\APACrefatitle {Adaptive evolution of honeybee dance dialects} {Adaptive evolution of honeybee dance dialects}.{\BBCQ}
\newblock
\APACjournalVolNumPages{Proceedings of the Royal Society B}{287}{1922}{20200190}.
\newblock
\begin{APACrefDOI} \doi{10.1098/rspb.2020.0190} \end{APACrefDOI}
\PrintBackRefs{\CurrentBib}

\bibitem [\protect \citeauthoryear {%
Lazaridou%
\ \BBA {} Baroni%
}{%
Lazaridou%
\ \BBA {} Baroni%
}{%
{\protect \APACyear {2020}}%
}]{%
lazaridou2020emergent}
\APACinsertmetastar {%
lazaridou2020emergent}%
\begin{APACrefauthors}%
Lazaridou, A.%
\BCBT {}\ \BBA {} Baroni, M.%
\end{APACrefauthors}%
\unskip\
\newblock
\APACrefYearMonthDay{2020}{}{}.
\newblock
{\BBOQ}\APACrefatitle {Emergent Multi-Agent Communication in the Deep Learning Era} {Emergent multi-agent communication in the deep learning era}.{\BBCQ}
\newblock
\APACjournalVolNumPages{arXiv preprint arXiv:2006.02419}{}{}{}.
\PrintBackRefs{\CurrentBib}

\bibitem [\protect \citeauthoryear {%
Lazaridou%
, Peysakhovich%
\BCBL {}\ \BBA {} Baroni%
}{%
Lazaridou%
\ \protect \BOthers {.}}{%
{\protect \APACyear {2017}}%
}]{%
lazaridou2017multi}
\APACinsertmetastar {%
lazaridou2017multi}%
\begin{APACrefauthors}%
Lazaridou, A.%
, Peysakhovich, A.%
\BCBL {}\ \BBA {} Baroni, M.%
\end{APACrefauthors}%
\unskip\
\newblock
\APACrefYearMonthDay{2017}{}{}.
\newblock
{\BBOQ}\APACrefatitle {Multi-Agent Cooperation and the Emergence of (Natural) Language} {Multi-agent cooperation and the emergence of (natural) language}.{\BBCQ}
\newblock
\BIn{} \APACrefbtitle {International Conference on Learning Representations (ICLR).} {International conference on learning representations (iclr).}
\PrintBackRefs{\CurrentBib}

\bibitem [\protect \citeauthoryear {%
Lewis%
}{%
Lewis%
}{%
{\protect \APACyear {1969}}%
{\protect \APACexlab {{\protect \BCnt {1}}}}}]{%
lewis1969communication}
\APACinsertmetastar {%
lewis1969communication}%
\begin{APACrefauthors}%
Lewis, D.%
\end{APACrefauthors}%
\unskip\
\newblock
\APACrefYearMonthDay{1969{\protect \BCnt {1}}}{}{}.
\newblock
{\BBOQ}\APACrefatitle {Convention: A Philosophical Study} {Convention: A philosophical study}.{\BBCQ}
\newblock
\BIn{} (\BCHAP~IV).
\newblock
\APACaddressPublisher{Cambridge, MA}{Harvard University Press}.
\PrintBackRefs{\CurrentBib}

\bibitem [\protect \citeauthoryear {%
Lewis%
}{%
Lewis%
}{%
{\protect \APACyear {1969}}%
{\protect \APACexlab {{\protect \BCnt {2}}}}}]{%
lewis1969coordination}
\APACinsertmetastar {%
lewis1969coordination}%
\begin{APACrefauthors}%
Lewis, D.%
\end{APACrefauthors}%
\unskip\
\newblock
\APACrefYearMonthDay{1969{\protect \BCnt {2}}}{}{}.
\newblock
{\BBOQ}\APACrefatitle {Convention: A Philosophical Study} {Convention: A philosophical study}.{\BBCQ}
\newblock
\BIn{} (\BCHAP~I).
\newblock
\APACaddressPublisher{Cambridge, MA}{Harvard University Press}.
\PrintBackRefs{\CurrentBib}

\bibitem [\protect \citeauthoryear {%
Lian%
, Bisazza%
\BCBL {}\ \BBA {} Verhoef%
}{%
Lian%
\ \protect \BOthers {.}}{%
{\protect \APACyear {2023}}%
}]{%
lian2023communication}
\APACinsertmetastar {%
lian2023communication}%
\begin{APACrefauthors}%
Lian, Y.%
, Bisazza, A.%
\BCBL {}\ \BBA {} Verhoef, T.%
\end{APACrefauthors}%
\unskip\
\newblock
\APACrefYearMonthDay{2023}{}{}.
\newblock
{\BBOQ}\APACrefatitle {Communication Drives the Emergence of Language Universals in Neural Agents: Evidence from the Word-order/Case-marking Trade-off} {Communication drives the emergence of language universals in neural agents: Evidence from the word-order/case-marking trade-off}.{\BBCQ}
\newblock
\APACjournalVolNumPages{Transactions of the Association for Computational Linguistics}{11}{}{1033--1047}.
\PrintBackRefs{\CurrentBib}

\bibitem [\protect \citeauthoryear {%
Lo%
, Gloag%
, Anderson%
\BCBL {}\ \BBA {} Oldroyd%
}{%
Lo%
\ \protect \BOthers {.}}{%
{\protect \APACyear {2010}}%
}]{%
https://doi.org/10.1111/j.1365-3113.2009.00504.x}
\APACinsertmetastar {%
https://doi.org/10.1111/j.1365-3113.2009.00504.x}%
\begin{APACrefauthors}%
Lo, N.%
, Gloag, R\BPBI S.%
, Anderson, D\BPBI L.%
\BCBL {}\ \BBA {} Oldroyd, B\BPBI P.%
\end{APACrefauthors}%
\unskip\
\newblock
\APACrefYearMonthDay{2010}{}{}.
\newblock
{\BBOQ}\APACrefatitle {A molecular phylogeny of the genus {Apis} suggests that the Giant Honey Bee of the Philippines, {A. breviligula Maa}, and the {Plains Honey Bee} of southern {India}, {A. indica Fabricius}, are valid species} {A molecular phylogeny of the genus {Apis} suggests that the giant honey bee of the philippines, {A. breviligula Maa}, and the {Plains Honey Bee} of southern {India}, {A. indica Fabricius}, are valid species}.{\BBCQ}
\newblock
\APACjournalVolNumPages{Systematic Entomology}{35}{2}{226-233}.
\newblock
\begin{APACrefURL} \url{https://resjournals.onlinelibrary.wiley.com/doi/abs/10.1111/j.1365-3113.2009.00504.x} \end{APACrefURL}
\newblock
\begin{APACrefDOI} \doi{https://doi.org/10.1111/j.1365-3113.2009.00504.x} \end{APACrefDOI}
\PrintBackRefs{\CurrentBib}

\bibitem [\protect \citeauthoryear {%
Macedonia%
\ \BBA {} Evans%
}{%
Macedonia%
\ \BBA {} Evans%
}{%
{\protect \APACyear {1993}}%
}]{%
macedonia1993variation}
\APACinsertmetastar {%
macedonia1993variation}%
\begin{APACrefauthors}%
Macedonia, J\BPBI M.%
\BCBT {}\ \BBA {} Evans, C\BPBI S.%
\end{APACrefauthors}%
\unskip\
\newblock
\APACrefYearMonthDay{1993}{}{}.
\newblock
{\BBOQ}\APACrefatitle {Variation among mammalian alarm call systems and the problem of meaning in animal signals} {Variation among mammalian alarm call systems and the problem of meaning in animal signals}.{\BBCQ}
\newblock
\APACjournalVolNumPages{Ethology}{93}{3}{177--197}.
\newblock
\begin{APACrefDOI} \doi{10.1111/j.1439-0310.1993.tb00988.x} \end{APACrefDOI}
\PrintBackRefs{\CurrentBib}

\bibitem [\protect \citeauthoryear {%
Nowak%
\ \BBA {} Krakauer%
}{%
Nowak%
\ \BBA {} Krakauer%
}{%
{\protect \APACyear {1999}}%
}]{%
nowak1999evolution}
\APACinsertmetastar {%
nowak1999evolution}%
\begin{APACrefauthors}%
Nowak, M\BPBI A.%
\BCBT {}\ \BBA {} Krakauer, D\BPBI C.%
\end{APACrefauthors}%
\unskip\
\newblock
\APACrefYearMonthDay{1999}{}{}.
\newblock
{\BBOQ}\APACrefatitle {The Evolution of Language} {The evolution of language}.{\BBCQ}
\newblock
\APACjournalVolNumPages{Proceedings of the National Academy of Sciences}{96}{14}{8028--8033}.
\PrintBackRefs{\CurrentBib}

\bibitem [\protect \citeauthoryear {%
Portegys%
}{%
Portegys%
}{%
{\protect \APACyear {2020}}%
}]{%
portegys2020morphognostic}
\APACinsertmetastar {%
portegys2020morphognostic}%
\begin{APACrefauthors}%
Portegys, T\BPBI E.%
\end{APACrefauthors}%
\unskip\
\newblock
\APACrefYearMonthDay{2020}{}{}.
\newblock
{\BBOQ}\APACrefatitle {Morphognostic Honey Bees Communicating Nectar Location Through Dance Movements} {Morphognostic honey bees communicating nectar location through dance movements}.{\BBCQ}
\newblock
\APACjournalVolNumPages{bioRxiv}{}{}{}.
\newblock
\begin{APACrefDOI} \doi{10.1101/2020.03.14.992263} \end{APACrefDOI}
\PrintBackRefs{\CurrentBib}

\bibitem [\protect \citeauthoryear {%
Schlenker%
\ \protect \BOthers {.}}{%
Schlenker%
\ \protect \BOthers {.}}{%
{\protect \APACyear {2026}}%
}]{%
schlenker2026ancestral}
\APACinsertmetastar {%
schlenker2026ancestral}%
\begin{APACrefauthors}%
Schlenker, P.%
, Lamberton, J.%
, Lan, N.%
, Lamberton, J.%
, Geraci, C.%
, Salis, A.%
\BDBL {}Chemla, E.%
\end{APACrefauthors}%
\unskip\
\newblock
\APACrefYearMonthDay{2026}{}{}.
\newblock
{\BBOQ}\APACrefatitle {Ancestral iconicity: the dance language of bees revisited} {Ancestral iconicity: the dance language of bees revisited}.{\BBCQ}
\newblock
\APACjournalVolNumPages{Biological Reviews}{101}{}{2033--2052}.
\newblock
\begin{APACrefDOI} \doi{10.1002/brv.70164} \end{APACrefDOI}
\PrintBackRefs{\CurrentBib}

\bibitem [\protect \citeauthoryear {%
Sch\"urch%
\ \BBA {} Gr\"uter%
}{%
Sch\"urch%
\ \BBA {} Gr\"uter%
}{%
{\protect \APACyear {2014}}%
}]{%
schurch2014dancing}
\APACinsertmetastar {%
schurch2014dancing}%
\begin{APACrefauthors}%
Sch\"urch, R.%
\BCBT {}\ \BBA {} Gr\"uter, C.%
\end{APACrefauthors}%
\unskip\
\newblock
\APACrefYearMonthDay{2014}{}{}.
\newblock
{\BBOQ}\APACrefatitle {Dancing Bees Improve Colony Foraging Success as Long-Term Benefits Outweigh Short-Term Costs} {Dancing bees improve colony foraging success as long-term benefits outweigh short-term costs}.{\BBCQ}
\newblock
\APACjournalVolNumPages{PLoS ONE}{9}{8}{e104660}.
\newblock
\begin{APACrefDOI} \doi{10.1371/journal.pone.0104660} \end{APACrefDOI}
\PrintBackRefs{\CurrentBib}

\bibitem [\protect \citeauthoryear {%
Seeley%
}{%
Seeley%
}{%
{\protect \APACyear {1989}}%
}]{%
seeley1989superorganism}
\APACinsertmetastar {%
seeley1989superorganism}%
\begin{APACrefauthors}%
Seeley, T\BPBI D.%
\end{APACrefauthors}%
\unskip\
\newblock
\APACrefYearMonthDay{1989}{}{}.
\newblock
{\BBOQ}\APACrefatitle {The Honey Bee Colony as a Superorganism} {The honey bee colony as a superorganism}.{\BBCQ}
\newblock
\APACjournalVolNumPages{American Scientist}{77}{6}{546--553}.
\PrintBackRefs{\CurrentBib}

\bibitem [\protect \citeauthoryear {%
Seeley%
\ \BBA {} Morse%
}{%
Seeley%
\ \BBA {} Morse%
}{%
{\protect \APACyear {1976}}%
}]{%
seeley1976nest}
\APACinsertmetastar {%
seeley1976nest}%
\begin{APACrefauthors}%
Seeley, T\BPBI D.%
\BCBT {}\ \BBA {} Morse, R\BPBI A.%
\end{APACrefauthors}%
\unskip\
\newblock
\APACrefYearMonthDay{1976}{}{}.
\newblock
{\BBOQ}\APACrefatitle {The nest of the honey bee ({Apis mellifera} L.)} {The nest of the honey bee ({Apis mellifera} l.)}.{\BBCQ}
\newblock
\APACjournalVolNumPages{Insectes Sociaux}{23}{4}{495--512}.
\newblock
\begin{APACrefDOI} \doi{10.1007/BF02223477} \end{APACrefDOI}
\PrintBackRefs{\CurrentBib}

\bibitem [\protect \citeauthoryear {%
Siregar%
, Le%
\BCBL {}\ \BBA {} Alhama%
}{%
Siregar%
\ \protect \BOthers {.}}{%
{\protect \APACyear {2026}}%
}]{%
siregar2026emergent}
\APACinsertmetastar {%
siregar2026emergent}%
\begin{APACrefauthors}%
Siregar, N\BPBI P.%
, Le, P.%
\BCBL {}\ \BBA {} Alhama, R\BPBI G.%
\end{APACrefauthors}%
\unskip\
\newblock
\APACrefYearMonthDay{2026}{}{}.
\newblock
{\BBOQ}\APACrefatitle {An emergent communication framework for honeybee waggle dance} {An emergent communication framework for honeybee waggle dance}.{\BBCQ}
\newblock
\BIn{} \APACrefbtitle {International Conference on the Evolution of Language} {International conference on the evolution of language}\ (\BPG~435).
\newblock
\begin{APACrefURL} \url{http://evolang.org/2026/proceedings/papers/evolang16_paper_94.pdf} \end{APACrefURL}
\PrintBackRefs{\CurrentBib}

\bibitem [\protect \citeauthoryear {%
Skyrms%
}{%
Skyrms%
}{%
{\protect \APACyear {2010}}%
}]{%
skyrms2010signals}
\APACinsertmetastar {%
skyrms2010signals}%
\begin{APACrefauthors}%
Skyrms, B.%
\end{APACrefauthors}%
\unskip\
\newblock
\APACrefYear{2010}.
\newblock
\APACrefbtitle {Signals: Evolution, Learning, and Information} {Signals: Evolution, learning, and information}.
\newblock
\APACaddressPublisher{Oxford}{Oxford University Press}.
\PrintBackRefs{\CurrentBib}

\bibitem [\protect \citeauthoryear {%
Smith%
, Kirby%
\BCBL {}\ \BBA {} Brighton%
}{%
Smith%
\ \protect \BOthers {.}}{%
{\protect \APACyear {2003}}%
}]{%
smith2003iterated}
\APACinsertmetastar {%
smith2003iterated}%
\begin{APACrefauthors}%
Smith, K.%
, Kirby, S.%
\BCBL {}\ \BBA {} Brighton, H.%
\end{APACrefauthors}%
\unskip\
\newblock
\APACrefYearMonthDay{2003}{}{}.
\newblock
{\BBOQ}\APACrefatitle {Iterated Learning: A Framework for the Emergence of Language} {Iterated learning: A framework for the emergence of language}.{\BBCQ}
\newblock
\APACjournalVolNumPages{Artificial Life}{9}{4}{371--386}.
\newblock
\begin{APACrefDOI} \doi{10.1162/106454603322694825} \end{APACrefDOI}
\PrintBackRefs{\CurrentBib}

\bibitem [\protect \citeauthoryear {%
Visscher%
\ \BBA {} Seeley%
}{%
Visscher%
\ \BBA {} Seeley%
}{%
{\protect \APACyear {1982}}%
}]{%
visscher1982foraging}
\APACinsertmetastar {%
visscher1982foraging}%
\begin{APACrefauthors}%
Visscher, P\BPBI K.%
\BCBT {}\ \BBA {} Seeley, T\BPBI D.%
\end{APACrefauthors}%
\unskip\
\newblock
\APACrefYearMonthDay{1982}{}{}.
\newblock
{\BBOQ}\APACrefatitle {Foraging Strategy of Honeybee Colonies in a Temperate Deciduous Forest} {Foraging strategy of honeybee colonies in a temperate deciduous forest}.{\BBCQ}
\newblock
\APACjournalVolNumPages{Ecology}{63}{6}{1790--1801}.
\newblock
\begin{APACrefDOI} \doi{10.2307/1940121} \end{APACrefDOI}
\PrintBackRefs{\CurrentBib}

\bibitem [\protect \citeauthoryear {%
von Frisch%
}{%
von Frisch%
}{%
{\protect \APACyear {1967}}%
}]{%
vonFrisch1967DanceLanguage}
\APACinsertmetastar {%
vonFrisch1967DanceLanguage}%
\begin{APACrefauthors}%
von Frisch, K.%
\end{APACrefauthors}%
\unskip\
\newblock
\APACrefYear{1967}.
\newblock
\APACrefbtitle {The Dance Language and Orientation of Bees} {The dance language and orientation of bees}\ (L\BPBI E.~Chadwick, \BTRANS{}).
\newblock
\APACaddressPublisher{Cambridge, MA}{Harvard University Press}.
\PrintBackRefs{\CurrentBib}

\bibitem [\protect \citeauthoryear {%
Wagner%
, Reggia%
, Uriagereka%
\BCBL {}\ \BBA {} Wilkinson%
}{%
Wagner%
\ \protect \BOthers {.}}{%
{\protect \APACyear {2003}}%
}]{%
wagner2003progress}
\APACinsertmetastar {%
wagner2003progress}%
\begin{APACrefauthors}%
Wagner, K.%
, Reggia, J\BPBI A.%
, Uriagereka, J.%
\BCBL {}\ \BBA {} Wilkinson, G\BPBI S.%
\end{APACrefauthors}%
\unskip\
\newblock
\APACrefYearMonthDay{2003}{}{}.
\newblock
{\BBOQ}\APACrefatitle {Progress in the Simulation of Emergent Communication and Language} {Progress in the simulation of emergent communication and language}.{\BBCQ}
\newblock
\APACjournalVolNumPages{Adaptive Behavior}{11}{1}{37--69}.
\PrintBackRefs{\CurrentBib}

\bibitem [\protect \citeauthoryear {%
Wheeler%
\ \BBA {} Fischer%
}{%
Wheeler%
\ \BBA {} Fischer%
}{%
{\protect \APACyear {2012}}%
}]{%
wheeler2012functionally}
\APACinsertmetastar {%
wheeler2012functionally}%
\begin{APACrefauthors}%
Wheeler, B\BPBI C.%
\BCBT {}\ \BBA {} Fischer, J.%
\end{APACrefauthors}%
\unskip\
\newblock
\APACrefYearMonthDay{2012}{}{}.
\newblock
{\BBOQ}\APACrefatitle {Functionally referential signals: A promising paradigm whose time has passed} {Functionally referential signals: A promising paradigm whose time has passed}.{\BBCQ}
\newblock
\APACjournalVolNumPages{Evolutionary Anthropology}{21}{5}{195--205}.
\newblock
\begin{APACrefDOI} \doi{10.1002/evan.21319} \end{APACrefDOI}
\PrintBackRefs{\CurrentBib}

\end{thebibliography}

\end{document}